\documentclass[english]{article}
\usepackage[T1]{fontenc}
\usepackage[latin9]{inputenc}
\usepackage{array}
\usepackage{float}
\usepackage{amsmath}
\usepackage{amssymb}
\usepackage{graphicx}
\usepackage{setspace}
\usepackage{esint}

\DeclareRobustCommand{\greektext}{%
  \fontencoding{LGR}\selectfont\def\encodingdefault{LGR}}
\DeclareRobustCommand{\textgreek}[1]{\leavevmode{\greektext #1}}
\DeclareFontEncoding{LGR}{}{}
\DeclareTextSymbol{\~}{LGR}{126}
\newcommand{\lyxmathsym}[1]{\ifmmode\begingroup\def\b@ld{bold}
  \text{\ifx\math@version\b@ld\bfseries\fi#1}\endgroup\else#1\fi}

\providecommand{\tabularnewline}{\\}

\makeatother

\usepackage{babel}
\begin{document}

\title{Preventing the tragedy of the commons through punishment of over-consumers
and encouragement of under-consumers}

\author{Irina Kareva$^{a,b,d}$, Benjamin Morin$^{a,b}$, Georgy Karev$^{c}$\\
 {\footnotesize $^{a}$ Mathematical, Computational Modeling Sciences
Center,}\\
 {\footnotesize {} PO Box 871904, Arizona State University, Tempe,
AZ 85287}\\
 {\footnotesize {} $^{b}$ School of Human Evolution and Social Changes,
}\\
 {\footnotesize {} Arizona State University, Tempe, AZ 85287 }\\
 {\footnotesize {} $^{c}$ Center for Biotechnology Information (NCBI)}\\
 {\footnotesize {} National Institutes of Health, Bethesda, 20892
}\\
 {\footnotesize {} $^{d}$ Center of Cancer Systems Biology}, {\footnotesize Steward
St. Elizabeth's Medical Center}\\
 {\footnotesize {} Tufts University School of Medicine, Boston, MA,
02135 }\\
 }
\maketitle
\begin{abstract}
\begin{doublespace}
The conditions that can lead to the exploitative depletion of a shared
resource, i.e, the tragedy of the commons, can be reformulated as
a game of prisoner's dilemma: while preserving the common resource
is in the best interest of the group, over-consumption is in the interest
of each particular individual at any given point in time. One way
to try and prevent the tragedy of the commons is through infliction
of punishment for over-consumption and/or encouraging under-consumption,
thus selecting against over-consumers. Here, the effectiveness of
various punishment functions in an evolving consumer-resource system
is evaluated within a framework of a parametrically heterogeneous
system of ordinary differential equations (ODEs). Conditions leading
to the possibility of sustainable coexistence with the common resource
for a subset of cases are identified analytically using adaptive dynamics;
the effects of punishment on heterogeneous populations with different
initial composition are evaluated using the Reduction theorem for
replicator equations. Obtained results suggest that one cannot prevent
the tragedy of the commons through rewarding of under-consumers alone
- there must also be an implementation of some degree of punishment
that increases in a non-linear fashion with respect to over-consumption
and which may vary depending on the initial distribution of clones
in the population.

\end{doublespace}

\end{abstract}

\section*{Introduction}

\begin{doublespace}
Ecological and social systems are complex and adaptive, composed of
diverse agents that are interconnected and interdependent \cite{miller2007complex,page2008difference,page2011diversity}.
Heterogeneity within these systems often drives the evolution and
adaptability of the system components enabling them to withstand and
recover from environmental perturbations. However, it is also heterogeneity
that makes the appearance and short-term prosperity of over-consumers
possible, which in turn can eventually lead to exhaustion of the common
resources (tragedy of the commons \cite{hardin2009tragedy}) and consequent
collapse of the entire population, also known as evolutionary suicide
\cite{j2005can}).

Elinor Ostrom has focused on the question of avoiding the tragedy
of the commons from the point of view of collective decision making
in small fisheries \cite{ostrom1990governing}. She observed that
the mutually satisfactory and functioning institution of collective
action could be developed in small communities where the possible
resource over-consumption by each individual was immediately noticeable
and punished, such as in water-monitoring systems instituted in Spanish
huertas, where each user was able to closely watch the other, and
where punishment for stealing scarce water was instituted not only
in monetary fines but also in public humiliation. This illustrates
a first path to prevent the tragedy of the commons: infliction of
punishment/tax for over-consumption, effectively selecting against
over-consumers. For instance, Ostrom cites as an example of a successful
adaptive governance system rural villages in Japan, where about 3
million hectares of forests and mountain meadows were successfully
managed even until now in no small part because ``accounts were kept
about who contributed to what to make sure no household evaded its
responsibilities unnoticed. Only illness, family tragedy, or the absence
of able-bodied adults ... were recognized as excuses for getting out
of collective labor... But if there was no acceptable excuse, punishment
was in order'' . Another situation, when the tragedy was successfully
avoided, is when the community introduces some kind of ``social currency'',
where an individual is rewarded for cooperation with social status
\cite{milinski2002reputation,vollan2010cooperation}. This is an example
of a second approach to preventing the tragedy of the commons: bestowing
reward/subsidy to under-consumers.

The dynamics of trade-offs between personal and population good have
been studied through classical game theory \cite{nowak2006evolutionary,vincent2005evolutionary,hofbauer2003evolutionary}.
However, many of the potentially relevant results have been obtained
for largely homogeneous populations of players. In this paper we approach
the question of preventing the tragedy of the commons through instituting
punishment/reward systems in heterogeneous populations using two recently
developed methods for studying evolving heterogeneous populations
in equation-based models, namely adaptive dynamics \cite{geritz1997evolutionarily}
and the Reduction theorem for replicator equations \cite{karev2010mathematical,karev2010principle}.

This paper is organized as follows: first we give a derivation of
a previously studied model, modified to incorporate the effects of
a punishment/reward system on overall population survival. Next, we
describe different approaches for studying the effects of punishment
and reward on population composition, namely, adaptive dynamics and
the Reduction theorem for replicator equations. Then we apply the
two techniques to a variety of punishment/reward structures, obtaining
both analytical and numerical results for sample populations that
differ in their initial composition with respect to over- and under-
consumers. We end the paper with a reformulation of the obtained results
in the context of the prisoner's dilemma, as well as with a comparison
of the two modeling methods. 
\end{doublespace}

\section*{Model Description}

\begin{doublespace}
In order to answer the questions posed in the introduction, we first
turn to a modified version of a mathematical model of `niche construction'
proposed by Krakauer et al.\cite{krakauer2009diversity}, where a
population of consumers $x(t)$ interact with a collectively constructed
`niche', represented as a common dynamical carrying capacity $z(t)$.
Rather than restricting the model to a specific physical resource,
we take the `niche' to be a generalized abstract `resource', such
as available nutrients, energy and other characteristics of the environment
that may affect both the individuals' survival and fecundity (in case
of financial systems, the 'resource' could be an inter-convertible
unit such as money) that each individual can invest or subtract from
the `common good'. Different individuals can contribute to the common
dynamical carrying capacity, or they can subtract from it at different
rates. Since sufficient reduction of the common `carrying capacity'
can indeed lead to population collapse, this model allows to model
effectively in a conceptual framework the question of the effects
of over-consumption on the survival of the population,

Within the frameworks of the model, each individual is characterized
by his or her own intrinsic value of resource consumption, denoted
by parameter $c$; a set of consumers that are characterized by the
same value of $c$ is referred to as $c$-clone. The total population
of all consumers is $N(t)=\sum_{\mathbb{A}}x_{c}$ if $c$ takes on
discrete values and $N(t)=\int_{\mathbb{A}}x_{c}dc$ if $c$ is continuous,
where $\mathbb{A}$ denotes the range of possible values of parameter
$c$.

The individuals $x_{c}(t)$ grow according to a logistic growth function,
where the carrying capacity is not constant but is a dynamic variable.
Each individual invests in resource restoration at a rate proportional
to $(1-c)$; an individual is considered to be an over-consumer if
$c>1$. Moreover, each individual is punished or rewarded according
to a general continuous function $f(c)\in C^{1}(\mathbb{\mathbb{R^{\dotplus}}})$
that directly affects the fitness of each consumer-producer such that
each clone is immediately punished for over-consumption if $c>1$
or rewarded for under-consumption when $c<1$.

The dynamics of the resource is determined by a natural restoration
rate $\gamma$ and decay/loss rate $\delta z(t)$. Restoration process
is accounted for by the term $e(1-c)\frac{N/z}{1+N/z}$, where $z(t)$
is restored both proportionally to each individual's investment, accounted
for by parameter $c$, and to the total amount of 'resource' consumed
by the entire population, represented by $N(t)/z(t)$.

This yields the following system of equations: 
\begin{eqnarray}
\underbrace{x'_{c}}_{\text{clones}}(t) & = & \underbrace{rx_{c}(t)}_{\text{intrinsic growth}}\left(\underbrace{c}_{\text{consumption}}-\underbrace{\frac{N(t)}{kz(t)}}_{\text{dynamical carrying capacity}}\right)+\underbrace{x_{c}(t)f(c)}_{\text{punishment/reward}},\label{eq:thesystem}\\
\underbrace{z'(t)}_{\text{resource}} & = & \underbrace{\gamma}_{\text{restoration}}+\underbrace{e\frac{\sum\nolimits _{\mathbb{A}}x_{c}(t)(1-c)}{z(t)+N(t)}}_{\text{change in resource caused by consumers}}-\underbrace{\delta z(t)}_{\text{decay/inflation}}.
\end{eqnarray}

The meaning of all the variables and parameters summarized in Table
\ref{tab: parameters meaning}.

\begin{table}[H]
\begin{tabular}{|c|c|c|}
\hline 
 & Meaning  & Range \tabularnewline
\hline 
\hline 
$z(t)$  & abstract generalized `resource'  & $z\geq0$ \tabularnewline
\hline 
$c$  & rate of resource consumption  & $c\in[c_{0},c_{f}]$ \tabularnewline
\hline 
$x_{c}(t)$  & population of individuals competing for the resource  & $x_{c}\geq0$ \tabularnewline
\hline 
$N(t)$  & total population size $N(t)=\intop_{\mathbb{\mathbb{A}}}x_{c}(t)dc$  & $N(t)\geq0$ \tabularnewline
\hline 
$r$  & individual proliferation rate  & $r\geq0$ \tabularnewline
\hline 
$k$  & `resource' conversion factor  & $k\geq0$ \tabularnewline
\hline 
$e$  & efficiency of construction of common `niche'  & $e\geq0$ \tabularnewline
\hline 
$\gamma$  & intrinsic rate of `resource' restoration independent of $x_{c}$  & $\gamma\geq0$\tabularnewline
\hline 
$\delta$  & per capita rate of natural `resource' decay  & $\delta\geq0$\tabularnewline
\hline 
$c_{0}$  & lower boundary value of parameter $c$  & $c_{0}\geq0$ \tabularnewline
\hline 
$c_{f}$  & upper boundary value of parameter $c$  & $c_{f}\geq0$ \tabularnewline
\hline 
\end{tabular}\caption{Summary of variables and parameters used in System \ref{eq:thesystem}.\label{tab: parameters meaning}}
\end{table}

\begin{table}[H]
{\center%
\begin{tabular}{|c|c|c|c|c|c|c|c|c|c|c|}
\hline 
 & $\gamma$  & $d$  & $e$  & $r$  & $k$  & $N_{0}$  & $z_{0}$  & $\mu$  & $c_{0}$  & $c_{f}$\tabularnewline
\hline 
\hline 
set 1  & 1  & 1  & 1  & 1  & 1  & 0.6  & 0.1  & 10  & 0  & 2.5\tabularnewline
\hline 
set 2  & 7.72  & 22  & 1  & 1  & 1  & 0.6  & 0.1  & 10  & 0  & 9.12\tabularnewline
\hline 
\end{tabular}\caption{Sample parameter values.\label{tab: Niche1 sample-parameter-values}}

} 
\end{table}

\begin{figure}
\centering{}\includegraphics[scale=0.5]{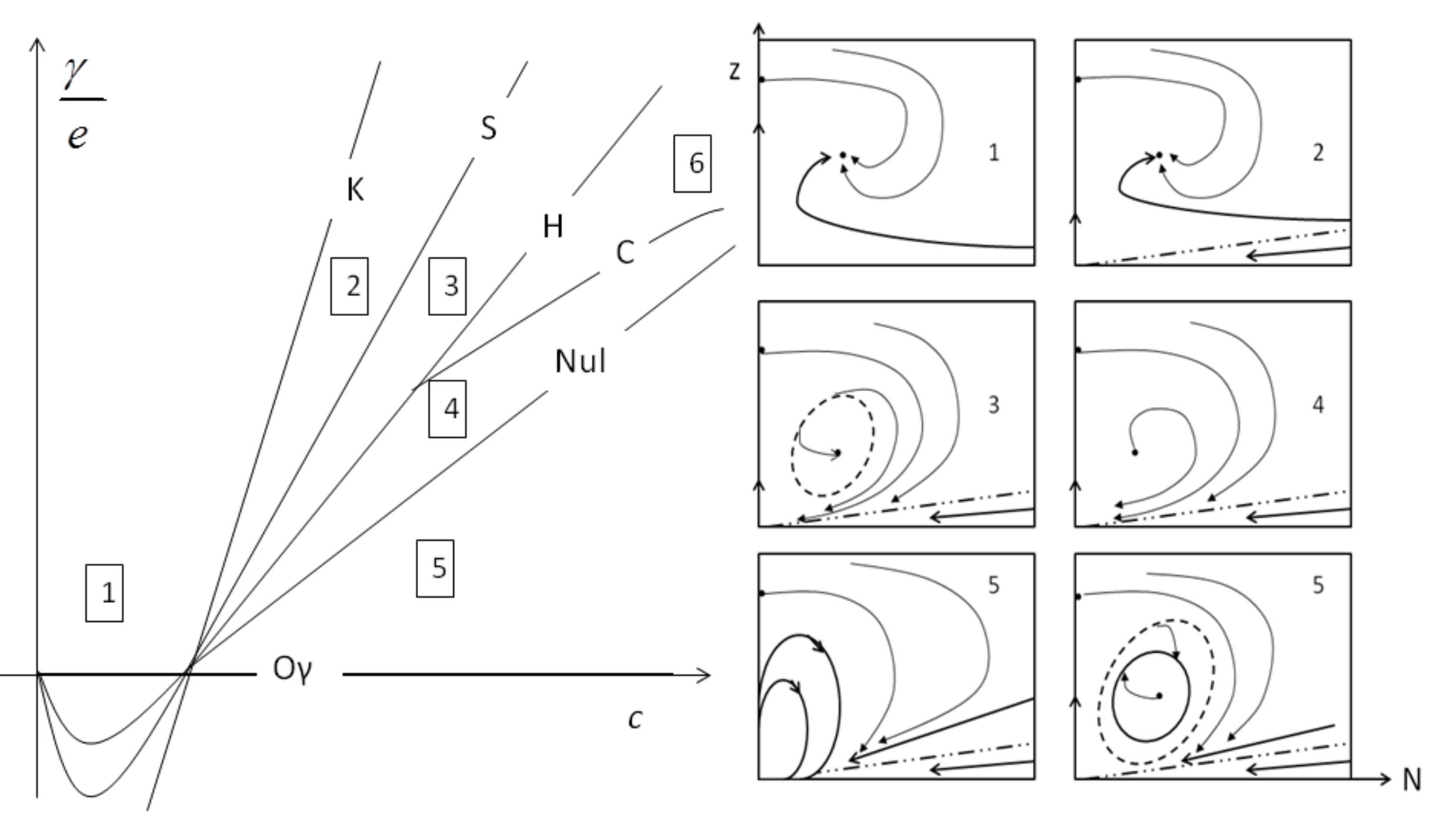}\caption{Bifurcation diagram of System (\ref{eq:thesystem}) when $f(c)=0$
in the $(\gamma,c)$ and \emph{(N,z)} phase spaces for fixed positive
parameters $e$ and $\delta$. The non-trivial equilibrium point $A$
is globally stable in Domain 1; it shares basins of attraction with
equilibrium $O$ in Domains 2 and 3. The separatrix of $O$ and the
unstable limit cycle that contains point $A$, serve, correspondingly,
as the boundaries of the basins of attraction. Only equilibrium $O$
is globally stable in Domains 4, which also contains also unstable
non-trivial $A$, and 5, which contains the elliptic sector. Domain
6 exists only for $\delta>5+\sqrt{24}$, where the stable limit cycle
that is in turn contained inside an unstable limit cycle, shares basins
of attraction with equilibrium $O$. Boundaries between Domains $K,S,H,Nul,C$
correspond respectively to appearance of an attractive sector in a
neighborhood of $O$, appearance of unstable limit cycle containing
$A$, change of stability of equilibrium $A$ via Hopf bifurcations,
disappearance of positive $A$ and saddle-node bifurcation of limit
cycles.\label{fig:Full-bifurcation-diagram}}
\end{figure}

The case, where $f(c)=0$ was previously completely studied in \cite{kareva2012transitional},
both for the case of a parametrically homogeneous and parametrically
heterogeneous system with respect to parameter $c$. Several transitional
regimes were identified with increasing parameter $c$, ranging from
sustainable coexistence with the common resource with ever decreasing
domain of attraction to sustained oscillatory regimes to collapse
due to exhaustion of the common resource. The results are summarized
in Figure \ref{fig:Full-bifurcation-diagram}. In Domain 1, when the
parameter of resource consumption is small, the common carrying capacity
remains large, successfully supporting the entire population, since
no individual is taking more resource than they restore. In Domain
2, a parabolic sector appears near the origin, decreasing the domain
of attraction of the non-trivial equilibrium point $A$. The population
can still sustainably coexist with the resource even with moderate
levels of over-consumption but the range of initial conditions, where
it is possible, decreases. As the value of $c$ is further increased,
the range of possible parameter values that allow sustainable coexistence
with the common resource decreases and is now bounded by the unstable
limit cycle, which appears around point $A$ through a catastrophic
Hopf bifurcation in Domain 3, and via generalized Hopf bifurcation
in Domain 6. Finally, in Domain 4 and 5, population extinction is
inevitable due to extremely high over-consumption rates unsupportable
by the resource.

In this paper we will investigate the question of whether timely infliction
of punishment for over-consumption or rewarding under-consumption
can prevent such a collapse and consequently prevent the tragedy of
the commons. We will also investigate the degree of effectiveness
of punishment depending on initial composition of the population with
respect to parameter $c$. Finally, we will try to investigate the
question of whether punishing those who over-consume or rewarding
those who under-consume will yield better results. These questions
will be addressed using two recently developed methods for modeling
parametrically heterogeneous populations, namely, adaptive dynamics
\cite{geritz1997evolutionarily} and Reduction theorem for replicator
equations \cite{karev2010mathematical,karev2010principle}. 
\end{doublespace}

\subsection*{Adaptive dynamics}

\begin{doublespace}
Adaptive dynamics is a series of techniques that have been recently
developed to address questions of system invasibility by rare `mutant'
clones. The main focus of this method is in evaluating a ``mutant's''
ability to proliferate in an environment set by the `resident' population
\cite{geritz1997evolutionarily}. We use this method to evaluate different
punishment/reward functions and their efficiency in preventing the
tragedy of the commons. It is worth noting that a more general theory
of systems with inheritance, which allows studying the problem of
`external stability' of a population of `residents' to invasion by
a `mutant' population, has been developed in \cite{gorban1984equilibrium};
it was later published in English in \cite{gorban2007selection}. 

Consider the equation for the dynamics of a rare mutant $x_{m}$ in
an environment set by the resident $x_{res}$. The total population
size is $x_{m}+x_{res}\approx x_{res}$, since $x_{m}$ is assumed
to be present at such low frequency that its contribution to the size
of the entire population is negligible.

Let $x_{res}^{*}$ satisfy $\frac{dx_{res}}{dt}=0$ , which implies
that $x_{res}^{*}=\frac{kz}{r}(f(c_{res})+rc_{res})$. Now introduce
a mutant, such that 
\begin{equation}
\frac{dx_{m}}{dt}=rx_{m}(c_{m}-\frac{x_{m}+x_{res}}{kz(t)})+x_{m}f(c_{m}).
\end{equation}

When the two subpopulations interact, the outcome of their interaction
is determined by the sign of the dominant eigenvalue of System 
\begin{equation}
\begin{split}\frac{dx_{res}}{dt} & =rx_{res}(c_{res}-\frac{x_{m}+x_{res}}{kz(t)})+x_{res}f(c_{res}),\\
\frac{dx_{m}}{dt} & =rx_{m}(c_{m}-\frac{x_{m}+x_{res}}{kz(t)})+x_{m}f(c_{m})\\
\frac{dz}{dt} & =\gamma+e\frac{x_{res}(1-c_{res})+x_{m}(1-c_{m})}{z+x_{m}+x_{res}}-\delta z
\end{split}
\label{eq:with mutant}
\end{equation}

The stability of the single positive non-trivial equilibrium point
$(x_{r}^{*},x_{m}^{*},z^{*})$ under the assumption that the mutant
is rare, i.e., when $x_{m}^{*}\approx0$, is determined by the dominant
eigenvalue of System \eqref{eq:with mutant}, which is given by $\lambda=r(c_{m}-c_{res})-f(c_{res})+f(c_{m})$:
the mutant will be able to invade if $\lambda>0$ and it will not
succeed if $\lambda<0$. Invasion fitness of the mutant, i.e., the
expected growth rate of a mutant in an environment set by the resident,
is given by 
\begin{equation}
\underset{T\rightarrow\infty}{lim}\frac{1}{T}\int_{0}^{T}\left[r(c_{m}-\frac{x_{res}(t)}{kz(t)})+f(c_{m})\right]dt=r(c_{m}-\frac{1}{k}\overline{(\frac{x_{res}}{z})})+f(c_{m})=r(c_{m},E_{res}),
\end{equation}
 where $\overline{(\frac{x_{res}}{z})}=\underset{T\rightarrow\infty}{lim}\frac{1}{T}\int_{0}^{T}\frac{x_{res}(t)}{kz(t)}dt$.
The selection gradient, which is defined as the slope of invasiogamen
fitness, determines if the invasion will be successful (positive sign
of the selection gradient predicts successful invasion) and is then
given by 
\begin{equation}
D(c_{m})=\frac{\partial}{\partial c_{m}}r(c_{m},E_{res})|_{[c_{m}=c_{res}]}=r+f'(c_{m}).\label{eq:selection gradient}
\end{equation}

These conditions allow answering a question of ``long-term'' invasibility,
i.e., whether a mutant with a slightly difference value of $c_{m}$
will be able to permanently invade the population of individuals,
characterized by parameter $c_{res}$. The points where selection
gradient is zero are known as evolutionarily singular strategies (ESS)
and are denoted here as $c_{res}^{*}$. Stability of $c_{res}^{*}$
for different types of punishment/reward functions is discussed below;
summary of some of the possible types of $c_{res}^{*}$ is given in
Table \ref{tab: Punish/reward comparing methods}. 

\begin{table}[H]
\begin{tabular}{|>{\centering}p{0.8in}|>{\raggedright}p{2.5in}|>{\raggedright}p{2.5in}|}
\hline 
 & Adaptive dynamics  & Reduction theorem\tabularnewline
\hline 
\hline 
{\small Goal}  & \multicolumn{2}{>{\raggedright}p{5in}|}{Model evolution of parametrically heterogeneous populations}\tabularnewline
\hline 
{\small Assumptions}  & \multicolumn{2}{>{\raggedright}p{5in}|}{Clonal reproduction}\tabularnewline
\hline 
 & \multicolumn{2}{>{\raggedright}p{5in}|}{Separation of evolutionary and ecological time scales}\tabularnewline
\hline 
{\small Environment}  & \multicolumn{2}{>{\raggedright}p{5in}|}{Can be variable and affected by changes in population composition
(also known as seascape, or a dancing landscape) }\tabularnewline
\hline 
{\small Population}  & Two types: invader and resident  & Any number of types\tabularnewline
\hline 
 & Small initial frequency of the invader  & Initial population composition can be arbitrary\tabularnewline
\hline 
 & Can introduce new ``mutants''  & All types must initially be present in the population, even if at
a near-zero frequency\tabularnewline
\hline 
 & \multicolumn{2}{>{\raggedright}p{5in}|}{Requires ``Lotka-Volterra'' type growth rates ($x'=xF(t)$)}\tabularnewline
\hline 
{\small Purpose/}{\small \par}

{\small question}  & Invasion: can a mutant invade the resident population?  & Evolution of a parametrically heterogeneous system over time due to
natural selection\tabularnewline
\hline 
 & \multicolumn{2}{>{\raggedright}p{5in}|}{Uses both theoretical analysis (bifurcation theory) and numerical
solutions}\tabularnewline
\hline 
{\small Visual representation}  & Pairwise invasibility plot (PIP)  & Bifurcation diagram of the corresponding parametrically homogeneous
system\tabularnewline
\hline 
\end{tabular}

\caption{A comparison of adaptive dynamics and the Reduction theorem\label{tab: Punish/reward comparing methods}}
\end{table}

\end{doublespace}

\subsection*{Different types of punishment/reward functions}

\subsubsection*{Case 1. Moderate punishment}

\begin{doublespace}
Consider the case, when the punishment function is of the form $f(c)=a\frac{1-c}{1+c}$.
This functional form allows to incorporate moderate levels of punishment/reward,
the severity of which is determined by the value of the parameter
$a$.

\begin{figure}[H]
 \centering{}\includegraphics[scale=0.5]{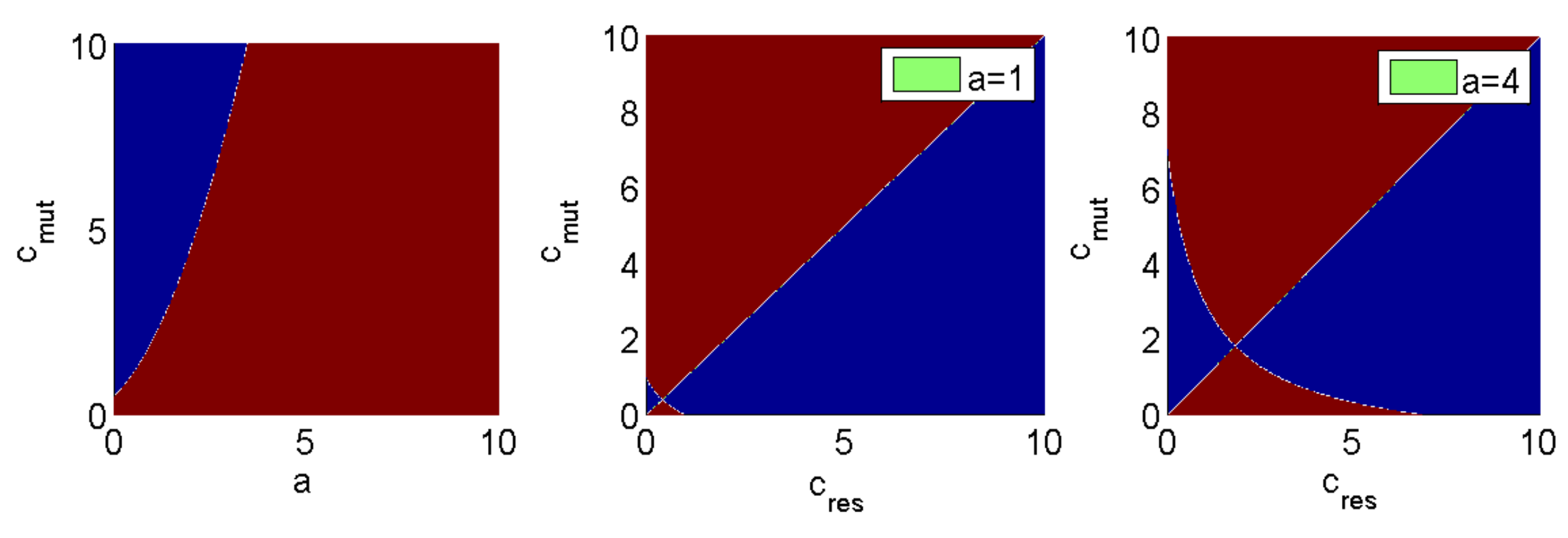}\caption{Selection gradient and pairwise invasibility plots (PIPs) for function
of the type $f(c)=a\frac{1-c}{1+c}$. Blue regions correspond to parameter
values where the mutant cannot invade; red regions correspond to parameter
values where it can invade. (a) selection gradient, defined in Equation
\eqref{eq:selection gradient} (b) PIP for $a=1$; the singular strategy
$c_{res}^{*}$ is evolutionarily unstable and convergence stable (c)
PIP for $a=4$; the singular strategy $c_{res}^{*}$ is evolutionarily
unstable and convergence stable. This punishment/reward function is
not effective against aggressive over-consumers. \label{fig:toc1}}
\end{figure}

The pairwise invasibility plot (PIP), which allows to visualize under
what relative values of $c_{m}$ and $c_{res}$ the dominant eigenvalue
$\lambda$ changes its sign, can be seen on Figure \ref{fig:toc1}b,c,
for $a=1$ and $a=4$. Blue regions correspond to the case when $\lambda(c_{res},c_{m})<0$,
and consequently the mutant cannot invade; red regions correspond
to the case when $\lambda(c_{res},c_{m})>0$, and the mutant can invade.
The point of intersection of the two curves corresponds to a convergence
stable (CSS) but evolutionarily unstable strategy, which can be invaded
by ``mutants'' with large enough values of $c_{m}$.

The selection gradient for this punishment function, $D(c_{m})=r-\frac{2a}{(1+c_{m})^{2}}$,
so the mutant with $c_{m}\approx c_{r}$ can invade when $a<\frac{r}{2}(1+c_{m})^{2}$
(see Figure \ref{fig:toc1} a).
\end{doublespace}

\begin{doublespace}
These results suggest that modest punishment can therefore protect
only from modest over-consumers. However, more aggressive over-consumers
cannot be kept of out the population, as the punishment is not severe
enough. 
\end{doublespace}

\subsubsection*{Case 2. Severe punishment/generous reward}

\begin{doublespace}
Now consider a case, when the punishment function is of the form $f(c)=a(1-c)^{3}$,
which allows to impose much more severe punishment on over-consumers
and more generous reward on under-consumers compared to the previous
case.

\begin{figure}[H]
 \centering{}\includegraphics[scale=0.5]{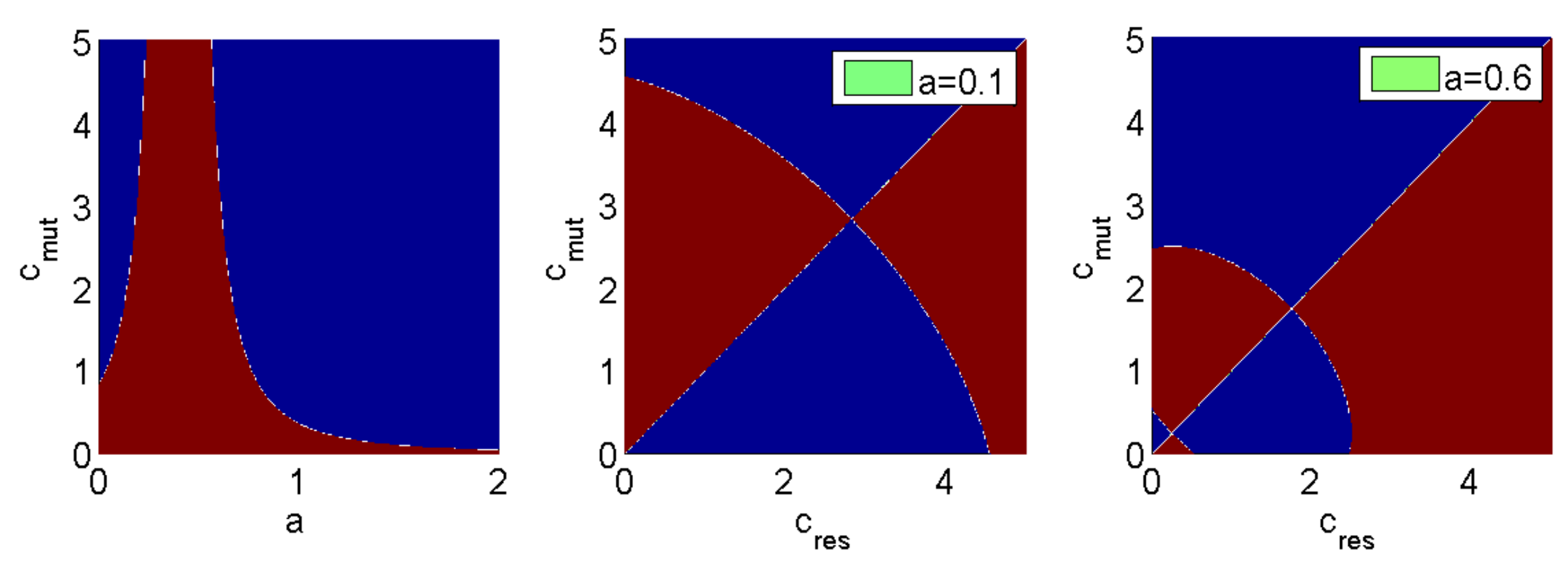}\caption{Selection gradient and pairwise invasibility plots for function of
the type $f(c)=a(1-c)^{3}$. Blue regions correspond to parameter
values where the mutant cannot invade; red regions correspond to parameter
values where it can invade. (a) selection gradient, defined in Equation
\eqref{eq:selection gradient} (b) PIP for $a=0.1$; the singular
strategy $c_{res}^{*}$ is evolutionarily and convergence stable (c)
PIP for $a=0.6$; there seem to appear two singular strategies for
large enough values of $a$; one of the strategies is evolutionarily
unstable, as neither the reward is large enough for the under-consumers,
nor the punishment severe enough for over-consumers; the second strategy
is $c_{res}^{*}$ is evolutionarily and convergence stable. This punishment/reward
function is effective against aggressive over-consumers.\label{fig: severe punish pip}}
\end{figure}

The PIP for this functional form can be seen on Figure \ref{fig: severe punish pip}
for $a=0.1$ and $a=0.6$. Once again, blue regions correspond to
the case when $\lambda(c_{res},c_{m})<0$, and consequently the mutant
cannot invade; red regions correspond to the case when $\lambda(c_{res},c_{m})>0$,
and the mutant can invade.

For this type of punishment/reward function, there is a region where
invader can invade but unlike the previous case, there exists an upper
boundary for the possible values of successful $c_{m}$. Unlike in
the previous case, singular strategy is stable, which predictably
implies that punishment needs to be severe enough in order to be able
to prevent invasion by over consumers. Moreover, for large enough
values of $a$, there can exist two singular strategies, one evolutionarily
stable and one evolutionarily unstable. In this case, the more ``altruistic''
ESS, which corresponds to smaller values of $c_{res}^{*}$, is unstable,
probably because in this case the reward for under-consumption is
insufficiently large, and the punishment is not sufficiently severe.
The second ESS, which corresponds to larger values of $c_{res}^{*}$
is evolutionarily stable, and so the mutant over-consumer cannot invade.

For this punishment function, the selection gradient is $D(c_{m})=r-3a(1-c_{m})^{2}$,
so $D(c_{m})>0$ if either $c_{m}>1-\sqrt{\frac{r}{3a}}$ or $c_{m}<1+\sqrt{\frac{r}{3a}}$.
One can see that there exists a region, where the mutant cannot invade
even when the enforcement of punishment, accounted for with parameter
$a$, is quite small. This can be interpreted as the rewards of over-consumption
being too small below the red invasibility zone, and the costs of
punishment being too great above the red invasibility zone. 

\end{doublespace}

\subsubsection*{Case 3. Separating punishment and reward}

\begin{doublespace}
Finally, consider the following punishment/reward function: $f(c)=\rho(1-c^{\eta})$.
This functional form allows to separate the impact of punishment for
over-consumption, which is increased or decreased depending on the
value of parameter $\eta$, and reward for under-consumption, which
is influenced primarily by the value of parameter $\rho$. In this
case, the dominant eigenvalue is given by $\lambda=r(c_{m}-c_{res})-\rho(c_{m}^{\eta}-c_{res}^{\eta})$.
As one can clearly see, if for instance $r=\rho$, then $\lambda>0$
when $\eta<1$, and so the mutant should be able to invade, since
the punishment for over-consumption is always less than its reproductive
benefits; if $\eta>1$, the inequality is reversed, and punishment
overwhelms reproductive benefits. If $\eta=1$, then everything is
determined solely by the relative values of growth rate $r$ and the
reward parameter $\rho$. If $c_{m}\approx c_{res}$, then the mutant
can (or cannot) invade if $D(c_{m})=r-\rho\eta c_{m}^{(\eta-1)}$
is positive (negative).

\begin{figure}[H]
 \centering{}\includegraphics[scale=0.5]{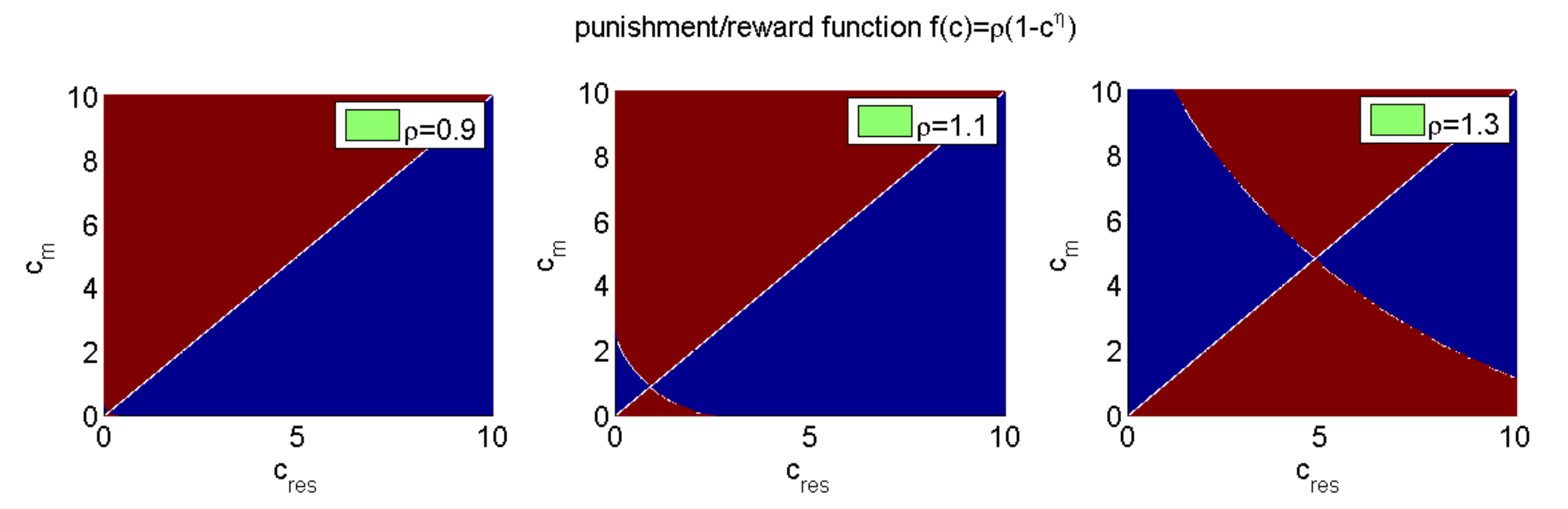}\caption{Pairwise invasibility plots for function type $f(c)=\rho(1-c^{\eta}),$
$\eta=0.9$. The effectiveness of this function is the same as was
for case 1: punishment is not severe enough to keep over-consumers
out of the population.\label{fig: case3 pip eta0_9}}
\end{figure}

\begin{figure}[H]
 \centering{}\includegraphics[scale=0.5]{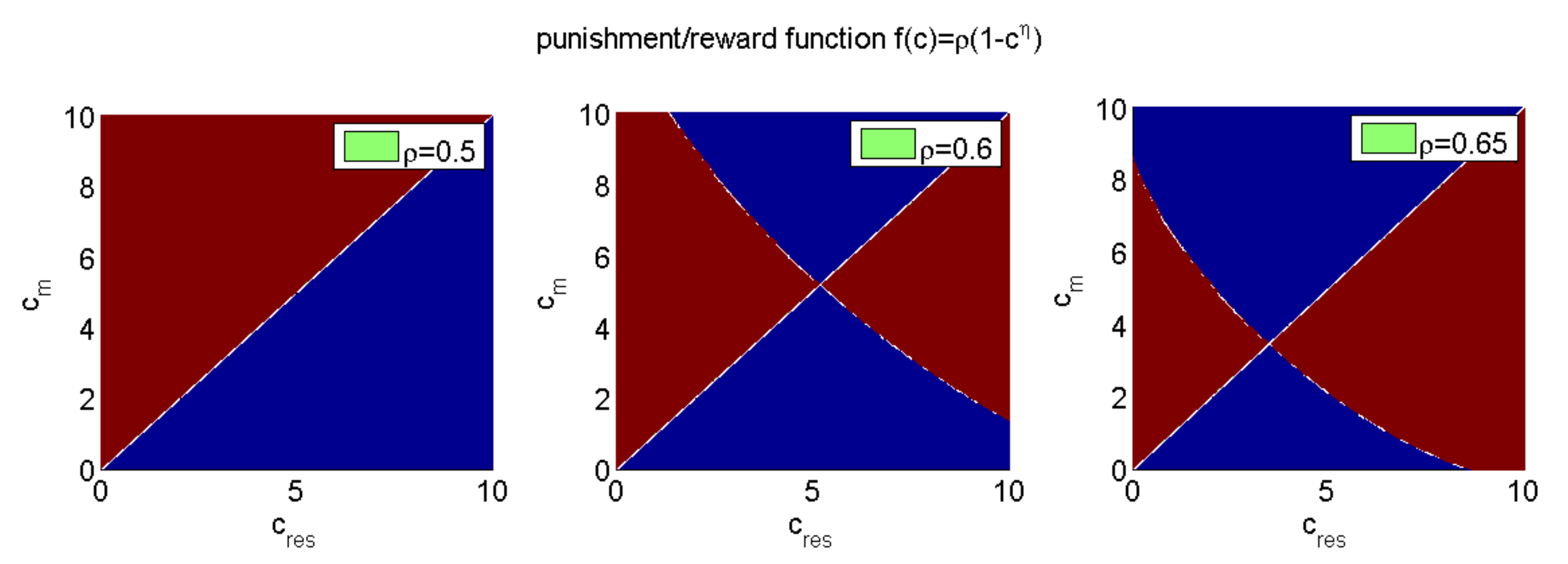}\caption{Pairwise invasibility plots for function type $f(c)=\rho(1-c^{\eta}),$
$\eta=1.2$. The effectiveness of this function is the same as was
for case 2: punishment is sufficiently severe to keep over-consumers
out of the population.\label{fig: case3 pip eta1_2}}
\end{figure}

As one can see on Figure \ref{fig: case3 pip eta0_9} and \ref{fig: case3 pip eta1_2},
this type of function behaves like case 1 for $\eta<1$ and like case
2 for $\eta>1$. These results reiterate the claim that was made in
the previous two cases: punishment needs to be severe enough in order
to successfully prevent invasion by over-consumers. 

\end{doublespace}

\subsection*{Modeling parametrically heterogeneous populations using the Reduction
theorem}

\begin{doublespace}
Adaptive dynamics allows to obtain analytical results about system
susceptibility to invasion in the case when the invading ``mutant''
(in this case an over-consumer with a higher value of $c_{m}$) is
rare. However, what if the initial mutant is not rare (invasion by
a group)? What if there is more than one type of mutant (the system
did not have time to recalibrate before the new mutant came along)?
We attempt to address these questions through application of the Reduction
theorem for replicator equations.

Without incorporating heterogeneity one cannot study the effects of
natural selection on the system, and until recently any attempts to
write such models resulted in systems of immense dimensionality. However,
the proposed approach allows us to overcome this problem.

Assume that each individual in the population studied is characterized
by its own value of some intrinsic parameter, such as birth, death,
resource consumption rates, etc. (a set of individuals characterized
by a particular value of the parameter studied is referred to here
as a clone). The distribution of clones will change over time due
to system dynamics, since different clones impose different selective
pressures on each other. Consequently, the mean of the parameter studied
also changes over time. The mean can be computed at each time point
using the moment generating function of the initial distribution of
clones in the population. The changes in the mean of the parameter
allows easy tracking of changes in population composition in response
to changes in micro-environment (such as varying nutrient availability)
or with respect to different initial distributions of clones.

Now let us introduce an auxiliary variable $q(t)$, which satisfies
$q(t)'=\frac{N(t)}{kz(t)}$, so that one may rewrite the system in
the following form:

\begin{align*}
\frac{x_{c}'(t)}{x_{c}(t)} & =r(c-q(t)')+f(c),\\
z'(t) & =p-dz(t)+e\frac{N(t)(1-E^{t}[c])}{N(t)+z(t)}.
\end{align*}

Then

\begin{equation}
x_{c}(t)=x_{c}(0)e^{-q(t)+t(rc+f(c))}.
\end{equation}

Total population size is then given by 
\begin{equation}
N(t)=\int_{c\in\mathbb{A}}x_{c}(t)dc=N(0)e^{-q(t)}\int_{c}e^{t(rc+f(c))}P_{c}(0)dc,
\end{equation}

where $P_{c}(0)=\frac{x_{c}(0)}{N(0)}$, and the current-time pdf
is given by 
\begin{equation}
P_{c}(t)=\frac{x_{c}(t)}{N(t)}=\frac{e^{t(rc+f(c))}P_{c}(0)}{\int e^{t(rc+f(c))}P_{c}(0)dc}.
\end{equation}

Now, the expected value of $c$ can be calculated from the definition:

\begin{equation}
E^{t}[c]=\int_{c}cP_{c}(t)=\frac{\int_{c}ce^{t(rc+f(c))}P_{c}(0)dc}{\int_{c}e^{t(rc+f(c))}P_{c}(0)dc}.
\end{equation}

The final system of equations becomes :

\begin{eqnarray}
z'(t) & = & p-dz+e\frac{N(t)(1-E^{t}[c])}{N(t)+z(t)},\\
q'(t) & = & \frac{N(t)}{kz(t)},
\end{eqnarray}
 where $N(t)$ and $E^{t}[c]$ are defined above.

The case, when $f(c)=0$ was investigated in \cite{kareva2012transitional}.
The authors observed that although it takes longer for a heterogeneous
population to go extinct, tragedy of the commons eventually happens
if the maximum value of $c$ is large enough. Moreover, one could
observe transitional regimes as the population was evolving towards
being composed of increasingly efficient over-consumers. This situation
can be used to forecast upcoming crisis and start implementing punishment
functions.

In the proposed form, if the system is parametrically homogeneous
system, i.e. when $E^{t}[c]$ is constant, $f(c)$ can be factored
into equation $\frac{x'}{x}=r(c+f(c)-\frac{bx}{kz})$, and consequently,
the spectrum of possible dynamical behaviors for this modified system
should be qualitatively the same compared to the initial model, analyzed
in \cite{kareva2012transitional} and summarized in Section 2. However,
while in \cite{kareva2012transitional} nothing prevented increase
of $E^{t}[c]$ up to the maximum possible value, here we want to investigate
punishment/reward functions that will prevent the increase of $E^{t}[c]$
that would otherwise drive the dynamics outside of the regions of
sustainable coexistence with the common resource. 
\end{doublespace}

\section*{Results}

\begin{figure}[H]
 \centering{}\includegraphics[scale=0.5]{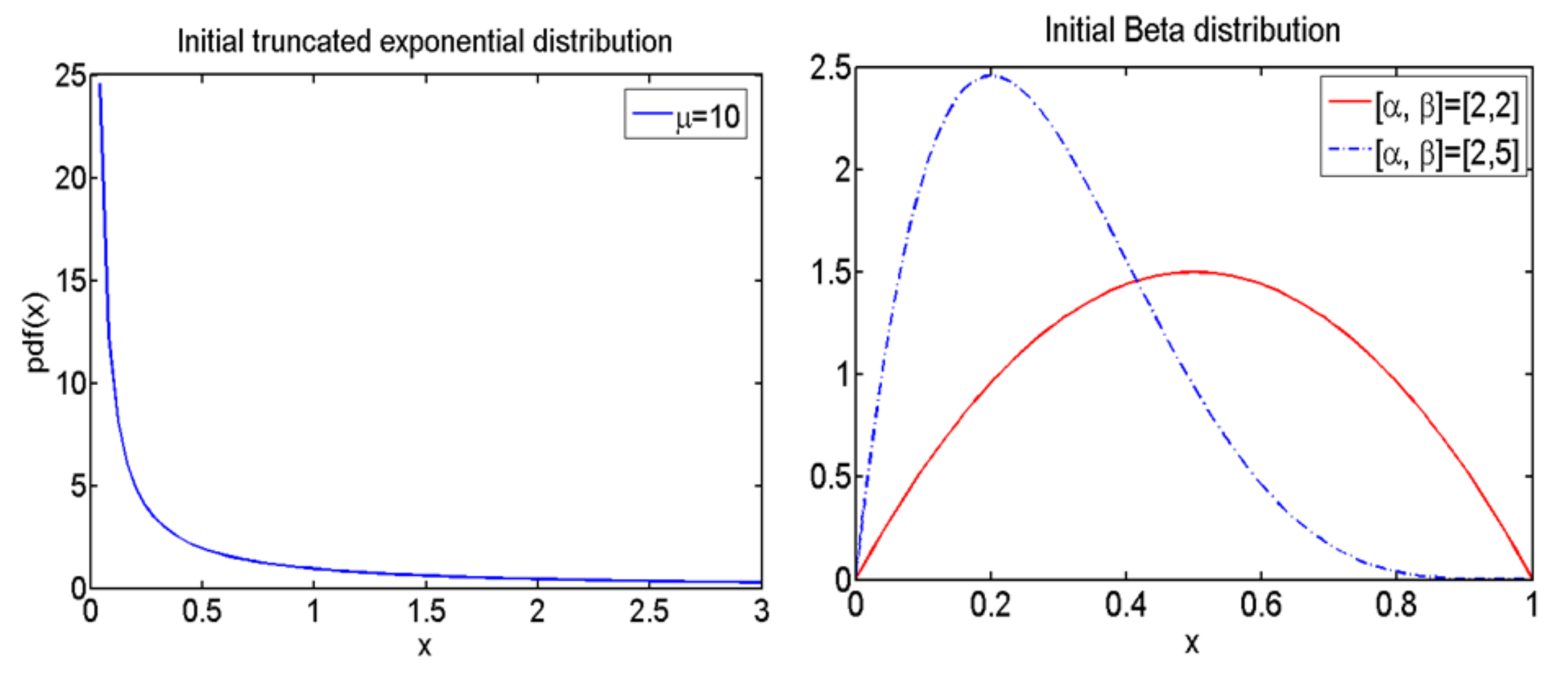}\caption{Initial distributions (a) truncated exponential with parameter $\mu=10$
(b) Beta distribution with parameters $\alpha=2,\beta=2$ and $\alpha=2,\beta=5$
.\label{fig:Initial-distributions}}
\end{figure}

\begin{doublespace}
We evaluated the effectiveness of several punishment/reward functions
on the dynamics of growth of a heterogeneous population of consumers
supported by a common dynamical carrying capacity, which can be increased
or decreased by the individuals themselves. We evaluated the effects
of the same type of punishment on populations with two different types
of initial distributions of clones, namely truncated exponential with
parameter $\mu=10$, and Beta distribution with parameters $\alpha=2\:,\beta=2$
and $\alpha=2,\:\beta=5$ (see Figure \ref{fig:Initial-distributions}).
Parameter values were chosen in such a way as to give different shapes
of the initial distribution of clones within the population We hypothesized
that the effectiveness of punishment will depend not only intrinsic
parameter values of the system but also on the initial composition
of the population. Specifically, we hypothesized that higher levels
of punishment/reward will be necessary for initial distributions,
where population composition is spread out further away from small
values of $c,$ such as Beta distributions with parameters $\alpha=2,\:\beta=5$
and even more so Beta distribution with parameters $\alpha=2,\:\beta=2$
(see Figure \ref{fig:Initial-distributions}b) compared to truncated
exponential distribution, where fewer over consumers are initially
present in the population (see Figure \ref{fig:Initial-distributions}a).

\begin{figure}[H]
\begin{centering}
\includegraphics[scale=0.65]{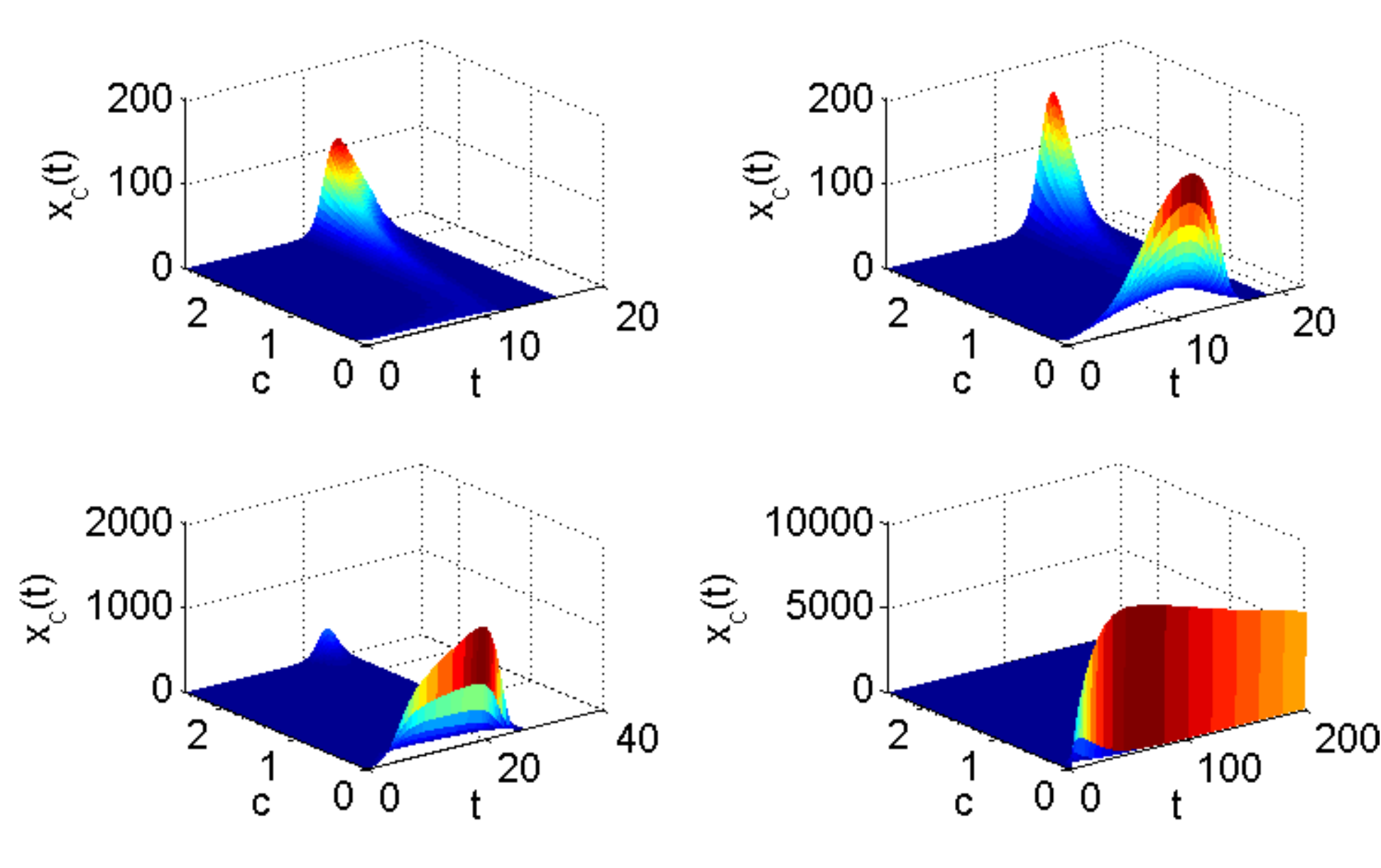} 
\par\end{centering}

\centering{}\caption{(a) Truncated exponential, moderate punishment (case 1), set 1, $a=0$.
(b) Truncated exponential, case 1, set 1, $a=0.5$. (c) Truncated
exponential, case 1, set 1, $a=1$. (d) Truncated exponential, case
1, set 1, $a=2$.\label{fig:Trexp s1c1 blue graphs}}
\end{figure}

\begin{figure}[H]
 \centering{}\includegraphics[scale=0.4]{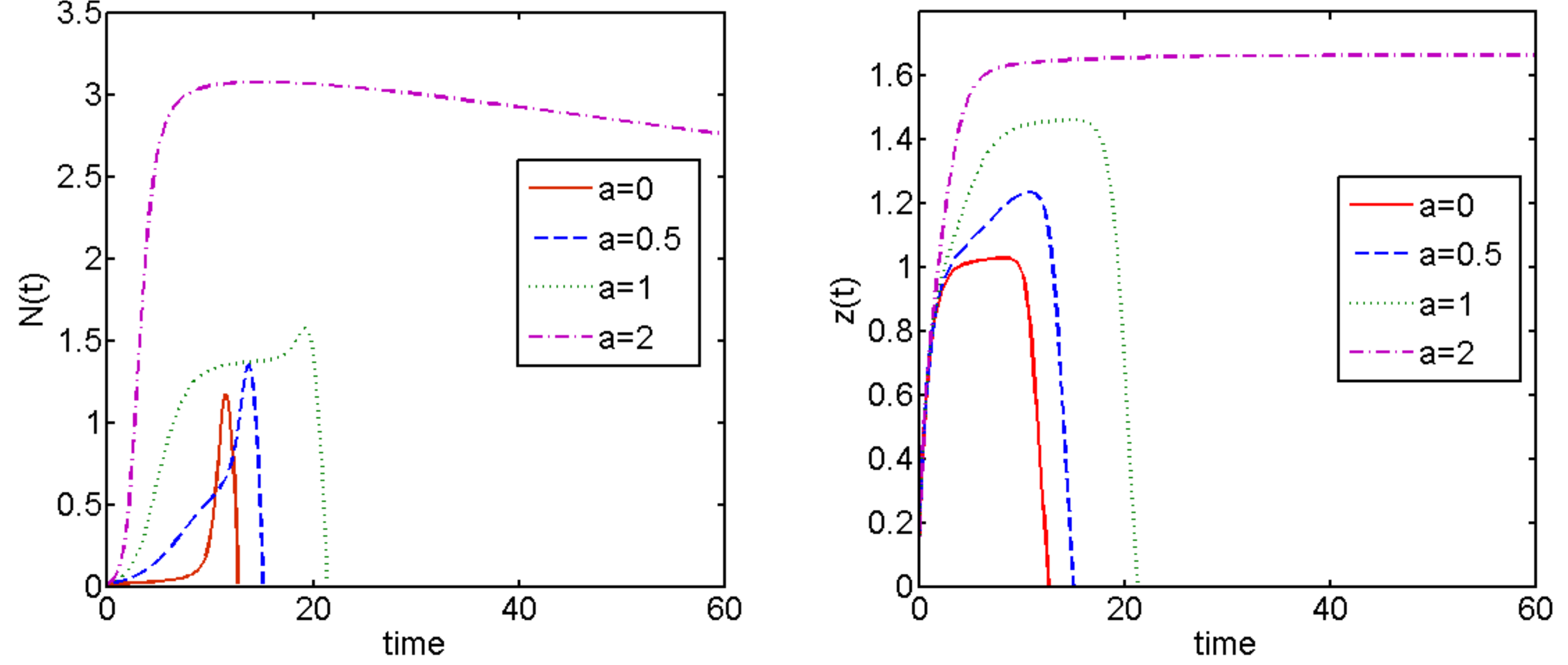}\caption{Truncated exponential distribution, set 1, case 1, dynamics of the
total population size and total resource with respect to different
values of $a$ (different levels of severity of imposed punishment).
One can see that successful management of over consumers was possible
only when punishment implementation was very high.\label{fig:Trexp s1c1 trajectories}}
\end{figure}

First, we evaluated the effectiveness of the moderate punishment function
of type $f(c)=a\frac{1-c}{1+c}$; the severity of punishment is captured
through varying parameter $a$. The initial distribution was taken
to be truncated exponential with parameter $\mu=10$. We took parameter
$a=0;\:0.5;\:1;2$ and plotted the changes in $x_{c}(t)$ for various
$c$ over time (Figure \ref{fig:Trexp s1c1 blue graphs}), as well
as the changes in the total population size and the amount of resource
(Figure \ref{fig:Trexp s1c1 trajectories}). We observed that when
the punishment imposed is moderate, over-consumption could be avoided
only when the value of $a$ was very high, i.e. when punishment is
imposed very severely.

Similar results were observed for the cases of Beta distribution with
parameters $\alpha=2,\:\beta=2$ and $\alpha=2,\:\beta=5$ (additional
figures are provided in Appendix). The value of $a$ that was necessary
for successful management of over-consumers varied depending on different
initial distributions, indicating that in order to be able to prevent
the tragedy of the commons, one needs to evaluate not only the type
of punishment and the severity of its enforcement but also match it
to the composition of the population, since one level of punishment
can be effective for one distribution of clones within a population
of consumers and not another. Noticeably, this kind of insight would
be impossible to obtain using the analytical methods of adaptive dynamics.

Next, we conducted the same set of numerical experiments for the severe
punishment\textbackslash{}generous reward function $f(c)=a(1-c)^{3}$.
We observed that the value of $a$ that would correspond to successful
restraint of over-consumers was much lower than in the previous case
for all initial distributions considered here (see Figures \ref{fig:beta22  c2s1 trajectories},
\ref{fig:beta22  c2s1 blue}; additional examples are provided in
Appendix). The system was able to support individuals with higher
values of parameter $c$ present in the initial population than in
the previous case. In some cases we were also able to observe brief
periods of oscillatory transitional dynamical behavior before the
system collapsed (see Figures \ref{fig:beta22  c2s1 blue} and \ref{fig:beta22  c2s1 trajectories}).

\begin{figure}[H]
 \centering{}\includegraphics[scale=0.65]{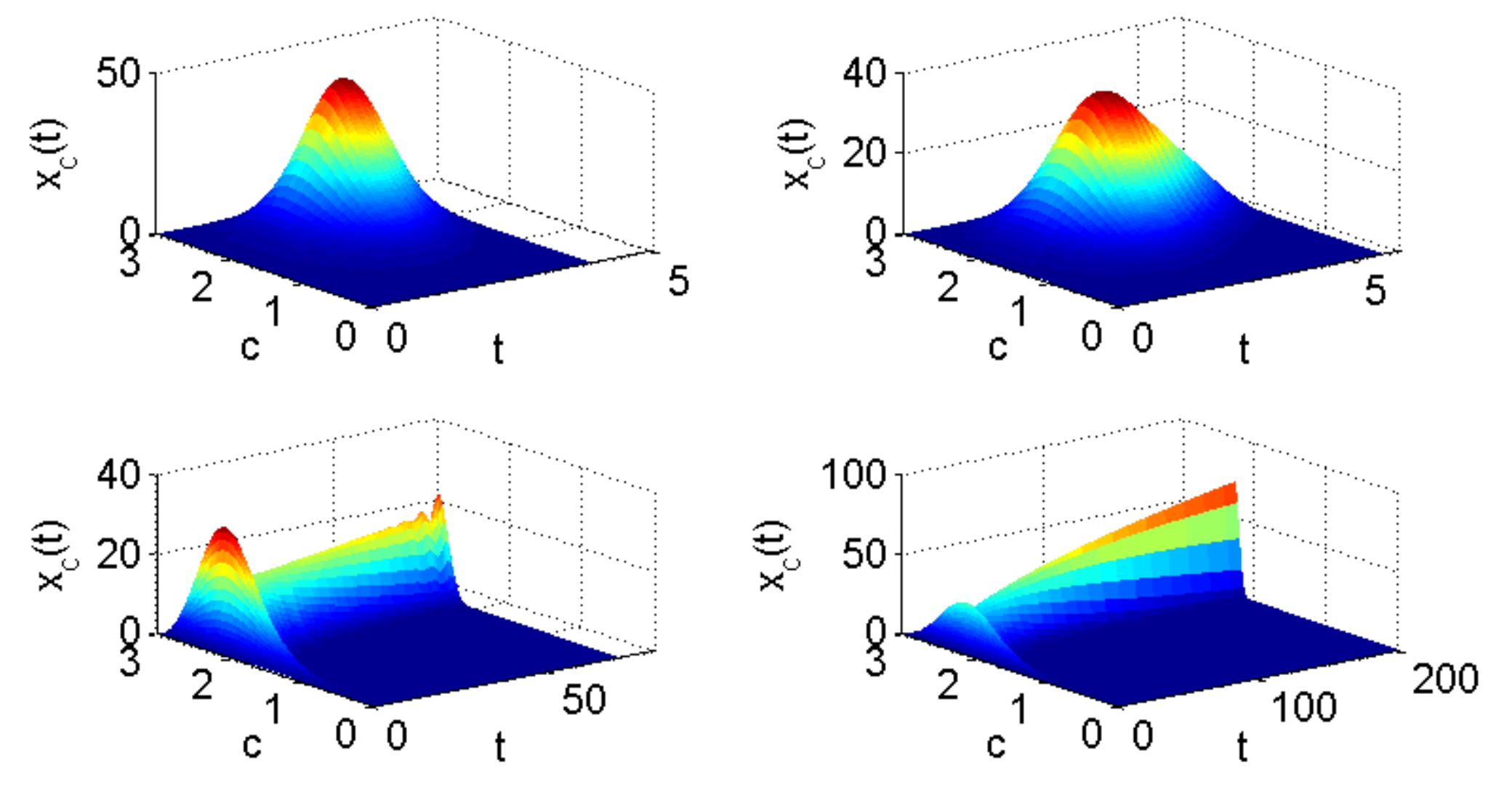}\caption{Severe punishment/generous reward (case 2). Beta distribution with
parameters $[\alpha,\beta]=[2,2]$. (a) $a=0$, (b) $a=0.1$, (c)
$a=0.17$, (d) $a=0.2$. One can see the population going through
transitional regimes before collapse, when the punishment is implemented
severely but not quite severely enough (when $a=0.7$). This most
probably corresponds to the expected value of parameter $c$ going
through region 3 in the phase parameter portrait of the non-distributed
system.\label{fig:beta22  c2s1 blue}}
\end{figure}

\begin{figure}[H]
 \centering{}\includegraphics[scale=0.4]{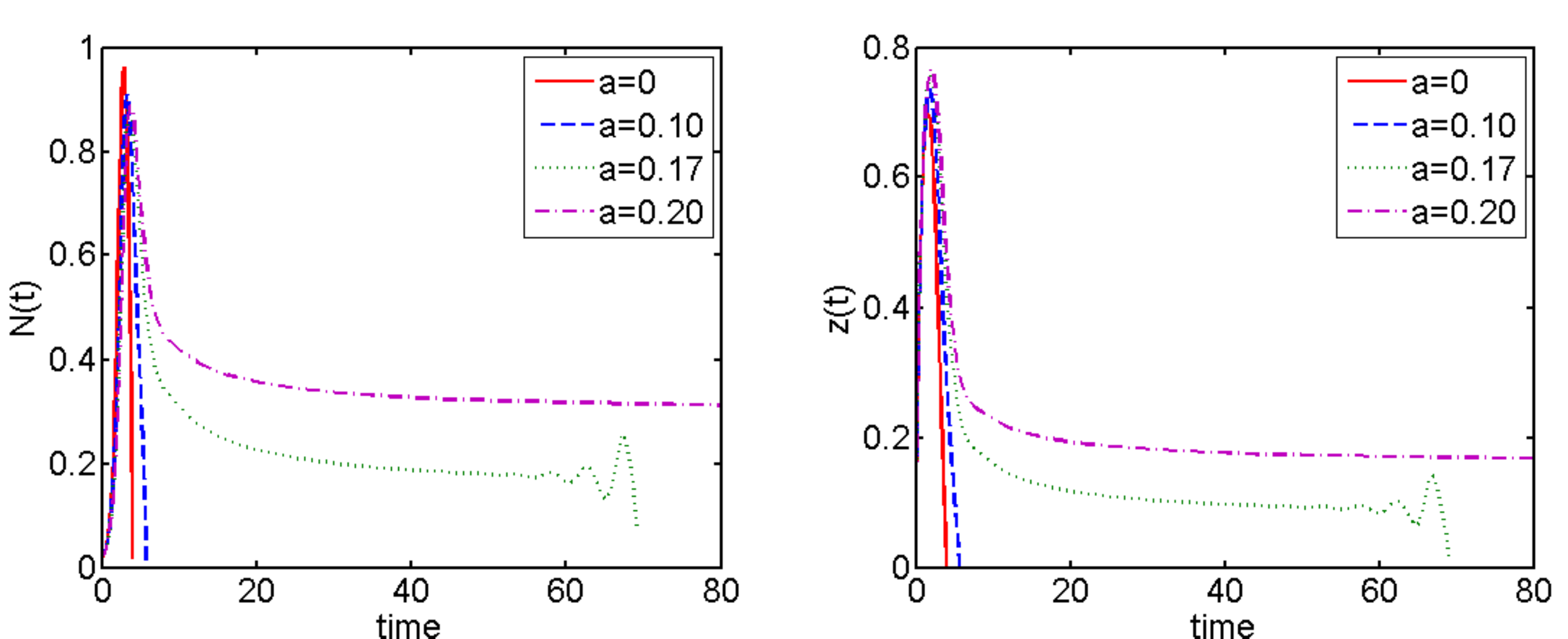}\caption{Case 2, set 1. Beta distribution with parameters $[2,2]$. Changes
in total population size and of the common resource over time.\label{fig:beta22  c2s1 trajectories}}
\end{figure}

\begin{figure}[H]
 \centering{}\includegraphics[scale=0.5]{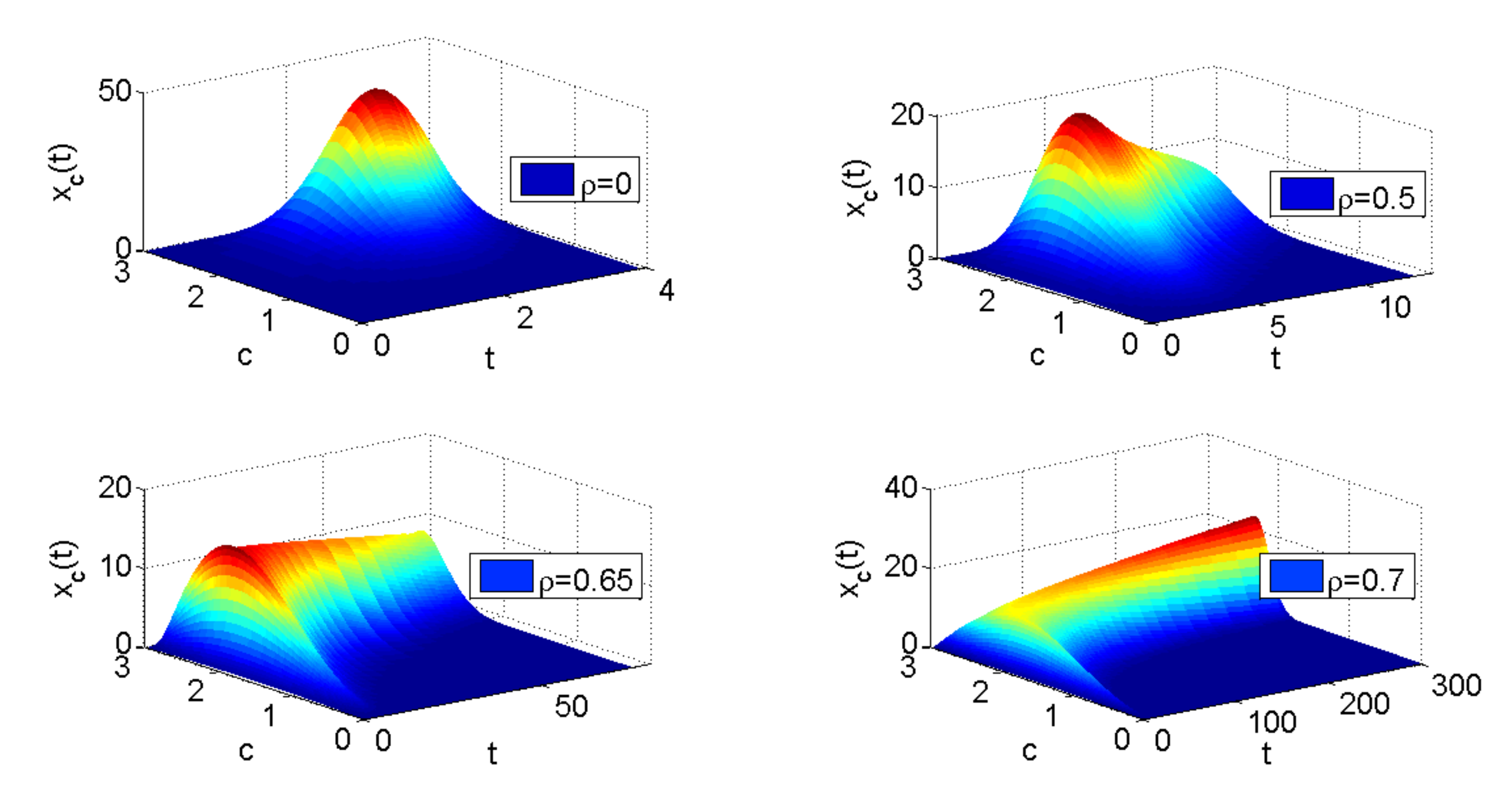}\caption{$f(c)=\rho(1-c^{\eta}),$ initial Beta distribution with parameters
$\alpha=2,\beta=2$, $c\in[0,3]$, $\eta=1.2$.\label{fig:case3 punish beta}}
\end{figure}

\begin{figure}[H]
 \centering{}\includegraphics[scale=0.5]{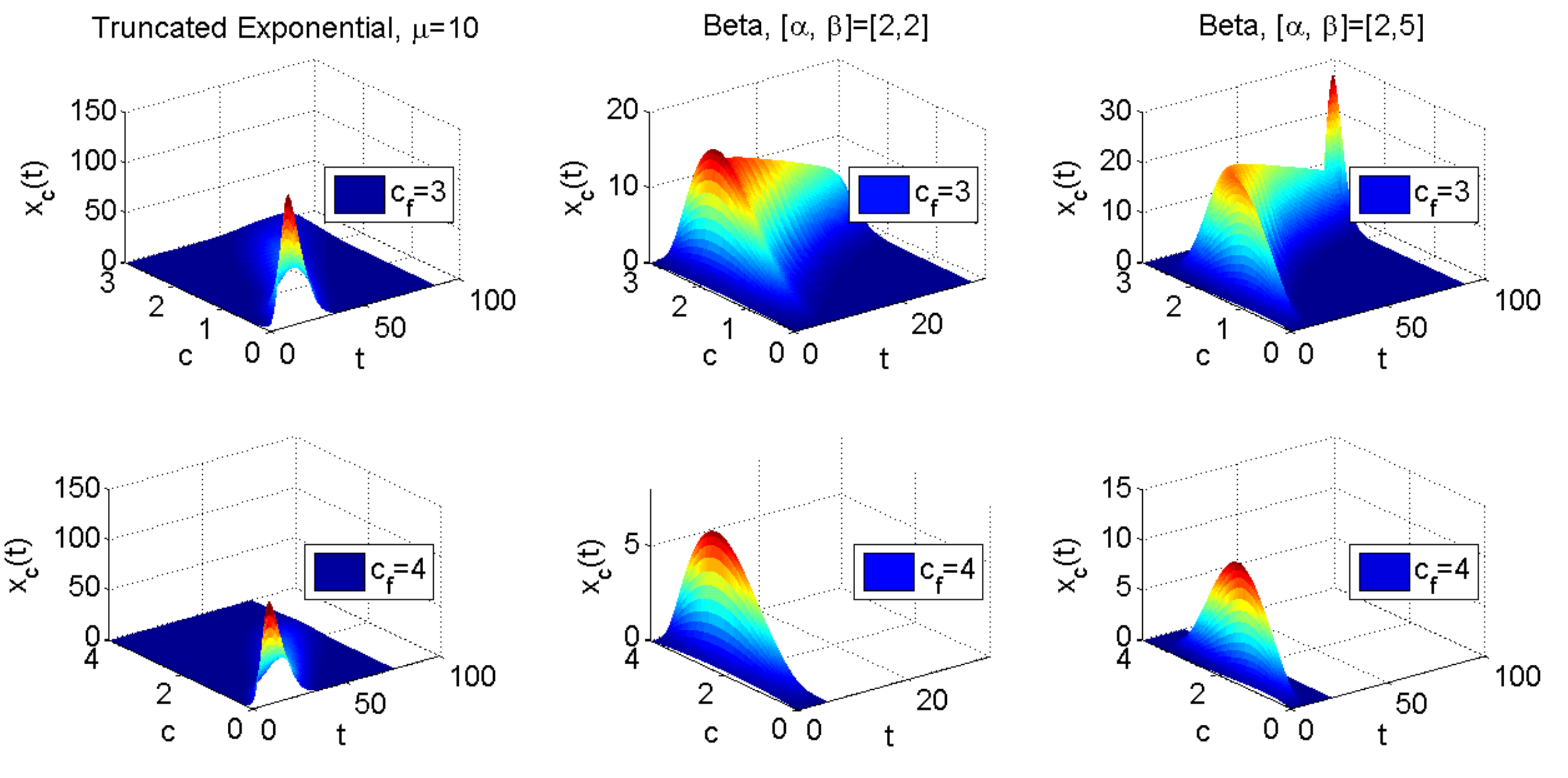}\caption{The importance of evaluating the range of possible values of $c_{f}$,
illustrated for different initial distribution. Punishment function
is of the type $f(c)=\rho(1-c^{\eta})$, where $\rho=0.6$, $\eta=1.2$.
Initial distributions are taken to be truncated exponential with parameter
$\mu=10$, and Beta with parameters $\alpha=2,\beta=2$ and $\alpha=2,\beta=5$;
$\rho=0.6$, $\eta=1.2$. The top row corresponds to $c\in[0,3]$;
the bottom row corresponds to $c\in[0,4]$.\label{fig:case 3 punish cf varied}}
\end{figure}

Finally, we evaluated punishment function of the type $f(c)=\rho(1-c^{\eta}),$
where the intensity of punishment and reward are accounted for by
parameters $\eta$ and $\rho$ respectively. We observed that in order
to evaluate the expected effectiveness of the punishment/reward system
one needs to not only adjust parameters $\rho$ and $\eta$ (see Figure
\ref{fig:case3 punish beta}) to each particular case considered but
also be able to evaluate the expected range of parameter $c$ (see
Figure \ref{fig:case 3 punish cf varied}), since one level of punishment
may be appropriate for one set of initial conditions but not another.
For instance, as one can see on Figure \ref{fig:case 3 punish cf varied},
the time to collapse under fixed values of parameter $\rho$ and $\eta$
is different for different initial distributions depending on the
maximum value of $c$ present in the initial population. Moreover,\textbf{
}in accordance with our hypothesis, indeed the time to collapse varies
depending on the initial distribution of the clones within the population,
and the higher the frequency of over-consumers is in the initial population,
the worse the prognosis. For the examples considered, population in
which the clones are distributed according to truncated exponential
distribution is less likely to collapse due to over-consumption than
Beta distribution with parameters $\alpha=2,\:\beta=5$ which in turn
is slightly less prone to collapse than Beta distribution with parameters
$\alpha=2,\:\beta=2$ \textbf{(}Figure \ref{fig:Initial-distributions}). 

\end{doublespace}

\subsection*{On the existence and properties of stationary distributions}

\begin{doublespace}
The adaptive dynamics approach assumes that there exists a stable
state of a population composed from a single clone $x_{res}$. In
the absence of punishment, any initial distribution evolves in such
a way, that asymptotically only the clone having maximal possible
value of the parameter $c$ will persist. In contrast, introducing
a punishment function allows for the possibility of a stationary distribution
of clones that is concentrated in more than one point. Specifically,
we show below that a population in a stationary state may consist
of up to 2 clones with punishment function of the 1st and 3rd type
and of up to 3 clones with punishment function of the 2nd type.

Assume that there exists a stationary distribution $P_{c}(t)=\frac{x_{c}(t)}{N(t)}$
of clones $x_{c}$ that stabilizes over time, which would occur when
\begin{equation}
\frac{dP_{c}(t)}{dt}=\frac{x_{c}'}{N}-\frac{x_{c}}{N^{2}}N'=P_{c}(t)[rc+f(c)-E^{t}[rc]-E^{t}[f(c)]]=0,\label{eq: change in ptc}
\end{equation}

which holds only when 
\begin{equation}
rc+f(c)=E^{t}[rc]+E^{t}[f(c)].\label{eq:condition for c}
\end{equation}

The left hand side of Equation \eqref{eq:condition for c} does not
depend on time, while the right hand side is independent of $c$,
which implies that Equation \eqref{eq:condition for c} holds only
when 
\begin{equation}
rc+f(c)=K,\label{eq:k fixed arb}
\end{equation}
 where $K$ is a constant.

Recall that $\{c_{i}\}$ is the support of a probability distribution
$P$ if $P(c)>0$ for all $c_{i}$ and $P(c)=0$ otherwise. In our
case, if there exists a constant $K$ such that 
\begin{equation}
f(c_{i})=K-rc_{i},\label{eq:14b}
\end{equation}
 then to each $K$ corresponds a stationary distribution $P(c_{i})$,
whose support $\{c_{i}\}$ coincides with the solution to this equation.
Therefore, Equation \eqref{eq:14b} can be used to identify the values
of $c$ that correspond to distributed evolutionary steady states.
(Note that a) any distribution concentrated in a single point $c$
is clearly stationary, and b) if $f(c)\equiv ac$, where $a$ is a
constant, then every stationary distribution is concentrated in a
single point.)

\begin{figure}[H]
 \includegraphics[scale=0.5]{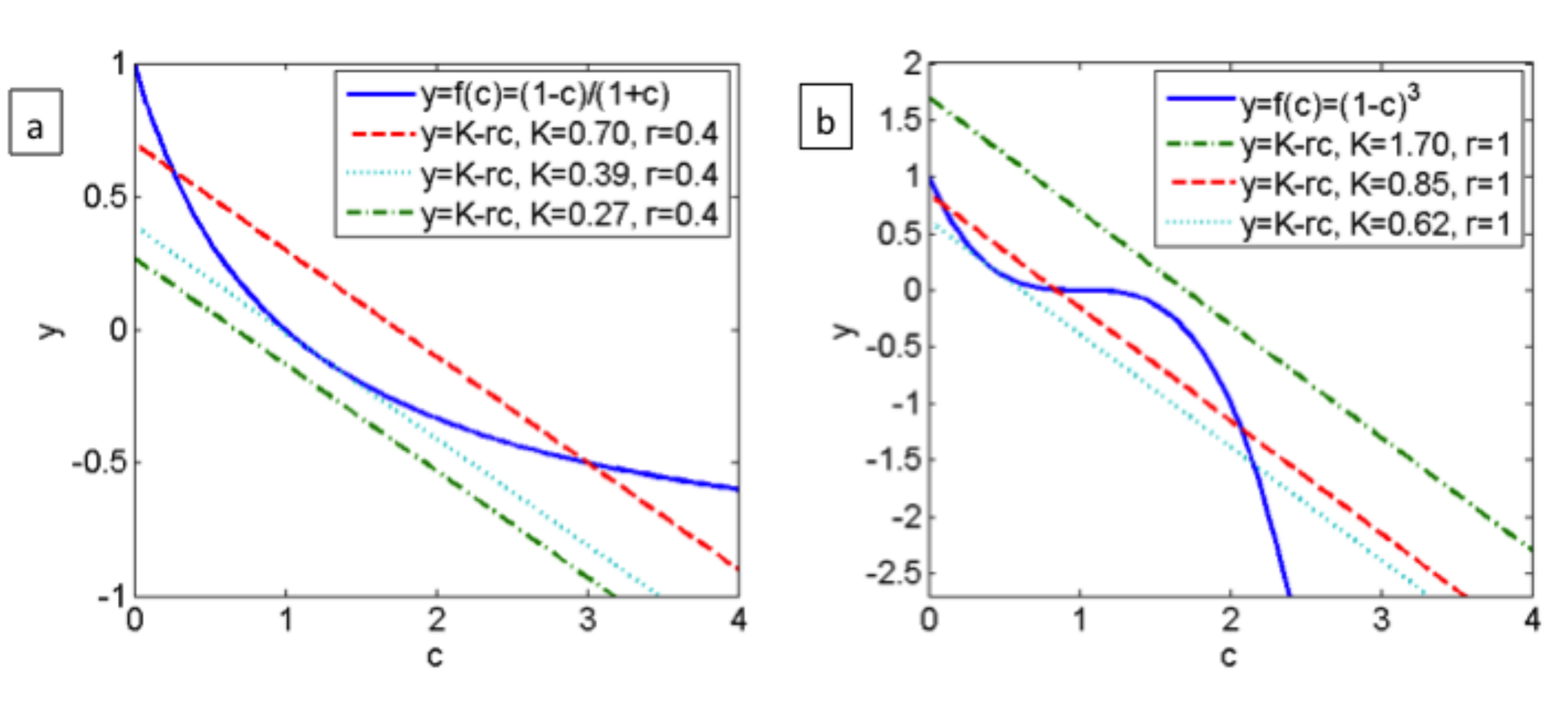}\caption{Steady states are concentrated at the points of intersection of punishment
function $y=f(c)$ and a line $y=K-rc$. Depending on the type of
punishment function and on the angle of the line $y=K-rc$, one may
have up to three clone types at the stationary state.\label{fig: stationary distributions}}
\end{figure}

It follows from Equation \eqref{eq:14b} that for a given punishment
function $f(c)$, the set of solutions to Equation \eqref{eq:14b}
at fixed $K$ can be represented geometrically as a set $S(K)$ of
abscissas of points of intersection of the curve $y=f(c)$ and the
line $y=K-rc$. The constant $K$ is free, so by changing K in such
a way that the set $S(K)$ is not empty, we obtain all possible supports
of the stationary distributions (see Figure \ref{fig: stationary distributions}).

An important conclusion can be made from these figures alone. As one
can see in Figure \ref{fig: stationary distributions}a, for instance,
the two points of intersection of $f(c)$ and $y=K-rc$ that correspond
to points of concentration of clones at a stationary state, can be
less than or greater than 1, depending on the parameter $r$, which
is the slope of the line $y=K-rc$. The case when one of the points
of intersection is less than one (``altruistic\textquotedblright{}
under-consuming clones) and one is greater than one (over-consumers)
can be interpreted as a population of under-consumers ``supporting\textquotedblright{}
the population of over-consumers at a stationary state. Asymptotic
behavior of the population, i.e., population survival or extinction,
will be determined by the dynamics of System \eqref{eq:14d}, which
is derived below, and corresponding parameter values.

Once again, let $\{c_{i}\}$ be a support of stationary probability
distribution. Then $\frac{dx_{c_{i}}}{dt}/x_{c_{i}}=r(K-\frac{N}{kz}).$
The right hand side does not depend on $i$, which implies that 
\[
\ln(x_{c_{i}}(t))-\ln(x_{c_{j}}(t))=const,\:\text{for all }i,j
\]
 and $\frac{x_{c_{i}}(t)}{x_{c_{j}}(t)}=\frac{x_{c_{i}}(0)}{x_{c_{j}}(0)}$
$\text{for all}\: t$. (Note that $i,\: j$ notation is used in this
section to distinguish the fact that we are no longer dealing with
rare clones in a resident population, and hence previously used $m,\: res$
notation is no longer appropriate.)

The dynamics of System \eqref{eq:with mutant} in this case can be
described by 
\begin{eqnarray}
\frac{dN}{dt} & = & rN(K-\frac{N}{kz})\label{eq:14d}\\
\frac{dz}{dt} & = & \gamma-\delta z+\frac{eN}{z+N}(1-E),\nonumber 
\end{eqnarray}
 where $E=E^{t}[c]=const$ and consequently does not depend on $t$.

Now let $A_{i}=\frac{x_{c_{i}}(0)}{\sum_{j}x_{c_{j}}(0)}$ be the
initial frequency of clones $x_{c_{i}}$; then $E=\sum_{i}A_{i}c_{i}$.
Therefore, by changing the initial values of $x(c_{i})$ we can vary
the value of $E$ from $min(c_{i})$ up to $max(c_{i})$. Stable equilibria
$(N^{*},\: z^{*})$ of System \eqref{eq:14d} can now be found from
$N^{*}=kKz^{*}$ and $\gamma-\delta z^{*}+\frac{\ensuremath{eN^{*}}}{z^{*}+N^{*}}(1-E)=\gamma-\delta z^{*}+B=0$,
where $B=\frac{(1-E)ekK}{1+ekK}=const$. Therefore, $z^{*}=\frac{\gamma+B}{\delta}$,
given that $\gamma+B>0$.

System \eqref{eq:14d} is very similar to the initial parametrically
homogeneous model with no punishment; the difference is that now the
values of $K$ and $E$ are not identical: the value $K$ is in one-to-one
correspondence with the support $\{c_{i}\}$, and the value of $E$
is defined by the support $\{c_{i}\}$ and initial values $\{x_{c_{i}}(0)\}$.
\emph{These results suggest that if the system has a stationary distribution,
then the dynamics of the resource would be determined not by absolute
sizes of the populations of `invaders' and `residents', but by their
ratio.}

Remark. In this work we are considering punishment functions, where
the support of the stationary distribution can consist of up to three
points, i.e., the system can be composed of up to three clones (see
Figure \ref{fig: stationary distributions}b). However, one can conceive
of a punishment function that can support an arbitrary number of clones
in the population at the stationary distribution, which correspond
to points of intersection of the curve $y=f(c)$ and the line $y=K-rc$.
Such system is of course unlikely to survive for any significant period
of time unless the number of individuals in the clones with $c^{*}\gg1$
is small enough. Such a case can be interpreted as a large number
of under-consumers supporting a very small number of over-consumers. 
\end{doublespace}

\subsubsection*{Example 1.}

\begin{doublespace}
Consider the punishment function of the form $f(c)=\frac{a(1-c)}{1+c}$;
the points of tangency of the curve $y=\frac{a(1-c)}{1+c}$ and the
line $y=K-rc$ are $c_{1,2}^{*}=\pm\sqrt{2a/r}-1$. In order for $c^{*}=\sqrt{2a/r}-1>0$,
condition $2a>r$ must be satisfied. At $c^{*}$, $f(c^{*})=\sqrt{2ar}-a$,
so the line $y=K-rc$ is tangent with $y=f(c)=\frac{a(1-c)}{1+c}$
at $K=K^{*}=\sqrt{2ar}-A+r(\sqrt{2a/r}-1)$ (see dotted line in Figure
\ref{fig: stationary distributions}a).

Hence, the line $y=K-rc$ intersects the curve $y=f(c)$ in two points
(see dashed red line in Figure\ref{fig: stationary distributions}a)
for all $K^{*}<K<a$.

For the purposes of illustration, take $a=1,\: r=0.4,\: K=0.7$ (see
Figure \ref{fig: stationary distributions}a). The stationary distribution
is concentrated in two points $c_{1}=0.25$ and $c_{2}=3$, so in
its steady state, the population is composed both of under-consumers
and of over-consumers. The population dynamics is then described by
system 
\begin{eqnarray*}
\frac{dN}{dt} & = & 0.4N(0.7-\frac{N}{kz}),\\
\frac{dz}{dt} & = & \text{\ensuremath{\gamma}}-\delta z+\frac{eN}{z+N}(1-E),
\end{eqnarray*}
 where $E\in[c_{1}=0.25,\: c_{2}=3]$. For these particular values
of parameters, the system will remain in the stable state if $\delta z^{*}=\lyxmathsym{\textgreek{g}}+\frac{eN^{*}}{z^{*}+N^{*}}(1-E)=\gamma+0.412e(1-E)>0$.
Otherwise, it will go to extinction. 
\end{doublespace}

\subsubsection*{Example 2.}

\begin{doublespace}
Now consider the punishment function of the form $f(c)=a(1-c)^{3}$.
The line $y=K-rc$ is tangent to the curve $y=f(c)$ at $c_{1,2}^{*}=1\pm\sqrt{r/(3a)}$.

Denote $\sqrt{r/(3a)}=\psi$. Then $c_{i,2}^{*}=1\pm\psi$; the condition
$\psi<1$ needs to be satisfied in order for both roots to be positive.

Consequently, 
\[
f(c_{1,2}^{*})=\mp\psi,
\]

\[
K_{1,2}^{*}=f(c_{2,1}^{*})+rc_{1,2}^{*}=r(1\pm\psi)\mp\psi^{3};
\]

and therefore 
\[
K_{min}=r(1+\psi)-\psi^{3}<K<r(1-\psi)+\psi^{3}=K_{max}.
\]

There can exist up to 3 points of intersection of the curve $f(c)=a(1-c)^{3}$
and the line $y=K-rc$ depending on the value of $K$.

As one can see in Figure \ref{fig: stationary distributions}b, where
$a=r=1,\:\psi=\sqrt{1/3},\: c_{1,2}^{*}=1\pm\sqrt{1/3}$, if $K>K_{max}=1.3849,$
then there are no solutions to Equation \eqref{eq:14b}; if $1<K<K_{max}$,
then there exist 2 solutions to Equation \eqref{eq:14b}; if $0.6151=K_{min}<K<1$,
then there can exist 3 solutions. For instance, taking $K=0.8$ and
solving equation $f(c)=K-rc,$ i.e., $(1-c)^{3}=0.8-c$, we obtain
the following solutions: $c_{1}=0.121,$ $c_{2}=0.791$, $c_{3}=2.08$.
So, in this case, at the steady state, two `altruistic' clones with
$c_{1}=0.121$ and $c_{2}=0.791$ are supporting over-consuming clones
with $c_{3}=2.088$.

The overall system dynamics for this example are given by 
\begin{eqnarray*}
\frac{dN}{dt} & = & rN(0.8-\frac{N}{kz}),\\
\frac{dz}{dt} & = & \text{\ensuremath{\gamma}}-\delta z+\frac{eN}{z+N}(1-E),
\end{eqnarray*}
 where $E\in[c_{1}=0.121,c_{3}=2.088].$ If $k=1$, then at the stationary
state $N^{*}=0.8z^{*}$ and $\frac{N^{*}}{z^{*}+N^{*}}=\frac{0.8}{1.8}=0.444$.
The system will remain at a steady state if 
\[
\delta z^{*}=\gamma+\frac{eN^{*}}{z^{*}+N^{*}}(1-E)=\gamma+0.444e(1-E)>0;
\]
 Otherwise, it will go to extinction. 
\end{doublespace}

\section*{Discussion}

\begin{doublespace}
In this paper we studied the dynamics of a consumer-resource type
system in order to answer the question of whether infliction of punishment
for over-consumption and reward for under-consumption can successfully
prevent, or at least delay, the onset of the tragedy of the commons.
We evaluated the effectiveness of three types of punishment functions,
as well as the effects of punishment on populations with different
initial composition of individuals with respect to the levels of resource
(over)consumption.

The proposed model was studied analytically in \cite{kareva2012transitional,krakauer2009diversity}
without incorporating punishment/reward for over-/under- consumption.
It describes the interactions of a population of consumers, characterized
by the value of an intrinsic parameter $c$, with a common renewable
resource in such a way that each individual can either contribute
to increasing the common dynamical carrying capacity ($c<1$) or take
more than they restore ($c>1$). As the value of $c$ increases, the
population goes through a series of transitional regimes from sustainable
coexistence with the resource to oscillatory regime to eventually
committing evolutionary suicide through decreasing the common carrying
capacity to a level that can no longer support the population.

We began by identifying analytical conditions leading to the possibility
of sustainable coexistence with the common resource for a subset of
cases using adaptive dynamics. This method allows to address the question
of whether a mutant (in our case, an individual with a higher value
of $c$) can invade a homogeneous resident population of resource
consumers. We evaluated the effectiveness of three types of punishment
functions: moderate punishment, $f(c)=a\frac{1-c}{1+c}$; severe punishment,
$f(c)=a(1-c)^{3}$, where the parameter $a$ denotes the severity
of implementation of punishment on individuals with the corresponding
value of parameter $c$, and function of the type $f(c)=\rho(1-c^{\eta})$,
which allows to separate the influence of reward for under-consumption,
primarily accounted for with parameter $\rho$, and punishment for
over-consumption, accounted for with parameter $\eta$.

We demonstrated that while moderate punishment/reward function can
be effective in keeping off moderate over-consumers at sufficiently
high values of $a$ (see Figures \ref{fig:Trexp s1c1 blue graphs}
and \ref{fig:Trexp s1c1 trajectories}), the evolutionarily singular
strategy in this case is unstable, and thus eventually the population
will be invaded by over-consumers with large enough value of $c$.
The severe punishment/generous reward approach was uniformly effective,
almost irrespective of the value of $a$, and allowed invasion of
moderate over consumers only in the small region of $c\approx1$,
when the punishment is not yet severe enough to outweigh the benefits
of moderate over-consumption (see Figures \ref{fig:beta22  c2s1 blue}
and \ref{fig:beta22  c2s1 trajectories}), and the reward does not
yet provide sufficient payoffs in terms of higher growth rates. The
evolutionarily singular strategy in this case is stable, suggesting
that punishment is severe enough to prevent the tragedy of the commons.

Finally, we investigated a punishment/reward function of the type
$f(c)=\rho(1-c^{\eta})$ that allowed separating the influence of
punishment for over-consumption (parameter $\eta$) from that of rewarding
under-consumption (parameter $\rho$). This functional type behaves
as a moderate punishment/reward function for $\eta<1$, rendering
it effective only for moderate over-consumers, and it behaves like
the severe punishment/reward function for $\eta>1$; in the critical
case $\eta=1$ the outcome of the interactions between resident and
invader populations is determined solely by relative values of growth
rate parameter $r$ and parameter $\rho$. Our results suggest that
just rewarding under-consumers is not enough to prevent invasion by
over consumers and hence one should not expect to be able to prevent
the tragedy of the commons through reward alone.

Adaptive dynamics techniques do not yet allow answering the question
of system invasibility when the mutant is not rare, such as in cases
of migration and consequent invasion by a group, or when the resident
population is inhomogeneous. We address these questions using the
Reduction theorem for replicator equations.

Assume that each individual consumer is characterized by his or her
own value of the intrinsic parameter $c$; a group of consumers with
this value of $c$ is referred to here as $c$-clones. This trait
directly affects fitness, and thus the distribution of clones will
change over time due to system dynamics. The clones will experience
selective pressures not only from the external environment, competing
for the limited resource, but also from each other. Consequently,
the mean of the parameter will also change over time, affecting system
dynamics. The mean of the parameter can be computed at each time point
from the moment generating function of the initial distribution of
clones, which allows to evaluate the effectiveness of different types
of punishment/reward functions on population composition by tracking
how the distribution of clones changed over time with respect to parameter
$c$.

We evaluated the effectiveness of the same three types of punishment/reward
functions on system evolution and calculated predicted the outcomes
for different initial distributions of clones within the population,
which were taken to be truncated exponential with parameter $\mu=10$,
and Beta distribution with parameters $\alpha=2,\:\beta=2$ and $\alpha=2,\:\beta=5$.
The initial distributions were chosen in such a way as to give significantly
different shapes of the initial probability density function; in applications
they should be matched to real data, when it is available. We observed
that the intensity of implementation of punishment\textbackslash{}reward
has to differ for different initial distributions if one is to successfully
stop over-consumption, and so in order to be able to make any reasonable
predictions one needs to understand what the initial composition of
the affected population is. We hypothesized that the higher the frequency
of over consumers in the initial population, the more severe the punishment
for over-consumption would have to be, and the more generous the reward;
specifically, for our examples, we anticipated the prognosis to be
the most favorable for initial truncated exponential distribution,
followed by Beta distribution with parameters $\alpha=2,\:\beta=5$
and then finally $\alpha=2,\:\beta=2$. This effect additionally implies
that the results obtained analytically from adaptive dynamics can
only be relevant for a subset of cases, i.e. when the invader is rare.

As anticipated, we observed that severe punishment/generous reward
approach was much more effective in preventing the tragedy of the
commons than the moderate punishment/reward function, particularly
for the cases, when over-consumers were present at higher frequencies
(such as both Beta initial distributions). Specifically, we observed
that the level of implementation, $a$, could be nearly ten times
lower for severe punishment/generous reward system as compared to
the moderate punishment function ($a\approx0.2$ vs $a\approx2$)
in order to obtain the same effect of selecting against the over-consumers,
which can be a very important factor in cases when there are large
costs associated with implementation of such intervention systems.
This comes not only from the severity of punishment but also from
the fact that moderate punishment allows more time for the over-consumers
to replicate, and thus by the time the punishment has an appreciable
effect, the population composition had changed, and the moderate punishment
will no longer be effective. So, in punishment implementation one
needs to take into account not only the severity of punishment but
also the time window that moderate punishment may provide, allowing
over-consumers to proliferate. Within the frameworks of the proposed
model, moderate implementation of more severe punishment\textbackslash{}reward
system is more effective than severe implementation of moderate punishment\textbackslash{}reward.

Proposed here is just one way to try and modify individuals' payoffs
in order to prevent resource over-consumption - through inflicting
punishment and reward that affects the growth rates of clones directly.
This approach can be modified depending on different situations, inflicting
punishment or reward based not just on the intrinsic value of $c$
but on total resource currently available. 
\end{doublespace}

\paragraph*{Adaptive dynamics and the Reduction theorem.}

\begin{doublespace}
In order to address our questions we have used two recently developed
methods for modeling parametrically heterogeneous populations: adaptive
dynamics \cite{geritz1997evolutionarily} and the Reduction theorem
for replicator equations \cite{karev2010mathematical,karev2010principle}.
Adaptive dynamics allows addressing the question of rare ``mutant''
clone invasion using standard bifurcation theory. The method allows
to formulate analytical conditions, which can be conveniently visualized
using pairwise invasibility plots (PIPs). However, the method does
not allow addressing questions of system invasibility by clones that
are not rare, such as in cases of invasion by a group, or when the
resident population is heterogeneous. These questions can be addressed
using the Reduction theorem for replicator equations.

We consider the moment of ``invasion'' as the initial time moment;
more formally, we assume that all ``invaders'' must be present initially
in the population, falling within some known distribution; one can
then see which clone type(s) will be favored over time due to natural
selection and visualize evolutionary trajectories, which cannot be
achieved using adaptive dynamics. The outcome depends both on the
initial distribution of the clones within the population and on the
initial state of the system, i.e., other intrinsic properties of both
individuals within the population and the resources, which in this
case were assumed to be fixed. However, the system of ODEs that results
from the transformation done using the Reduction theorem is typically
non-autonomous; hence no analytical conditions can typically be obtained
using standard bifurcation theory. The two methods therefore can complement
each other. A more detailed comparison of the two methods is in Table
\ref{tab: Punish/reward comparing methods}. 
\end{doublespace}

\begin{doublespace}

\subsubsection*{Tragedy of the commons as prisoner's dilemma}
\end{doublespace}

\begin{doublespace}
The conditions that can lead to the tragedy of the commons can be
reformulated as a game of prisoner's dilemma: in such systems preserving
the common resource is in the best interest of the group as a whole
but over-consumption is in the interest of each particular individual.
In the cases where decisions about the resource are made by each individual
independently, tragedy of the commons seems to be inevitable. However,
a number of cases have been observed, when the tragedy could be avoided
if the individuals within the population were able to communicate
with each other \cite{ostrom1990governing}. The common thread that
runs through many of the available examples is that 1) in small groups
the effects of over-consumption are immediately noticeable, since
within a small population each individual's actions are more visible
than in larger populations, and 2) punishment is enforced without
delay.

\begin{figure}
\centering{}\includegraphics[scale=0.55]{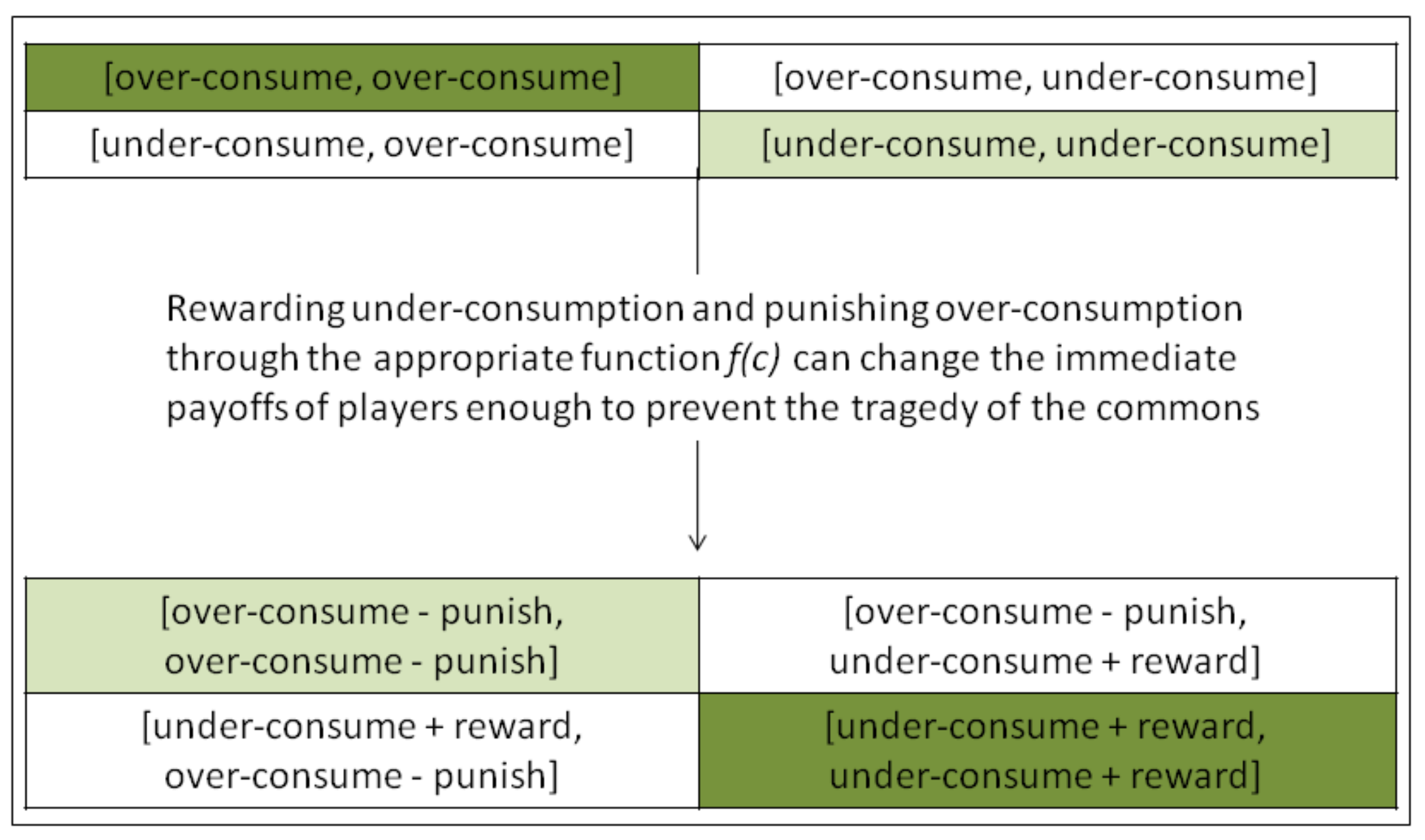}\caption{Tragedy of the commons as prisoner's dilemma. The tragedy can be avoided
if the immediate payoffs of all the players are modified through appropriate
punishment/reward functions.\label{fig:prisoner's dilemma matrix-1}}
\end{figure}

In this paper we are dealing with a particular case of a game of prisoner's
dilemma, where punishment/reward functions $f(c)$ affect individual
payoffs in such a way as to give the possibility to outweigh both
the benefits of over-consumption and the deleterious effects of under-consumption
(see Figure \ref{fig:prisoner's dilemma matrix-1}). We were able
to demonstrate that preventing the tragedy of the commons through
solely rewarding under-consumers is unlikely. However, the effectiveness
of ``intervention'' is increased when both reward and punishment
systems are enforced, although they do need to be modified depending
on population composition, since the choices of individuals may be
affected not only by their immediate `payoffs' but also by the actions
of those surrounding them. Therefore, if one is to try and avert the
tragedy of the commons through punishing over-consumption and rewarding
under-consumption, one needs to not only find an appropriate punishment/reward
function but also calibrate it to the current composition of each
particular population. 

\end{doublespace}

\section*{Acknowledgments }

\begin{doublespace}
The authors wougameld like to thank Kalle Parvinen and John Nagy for
their help with adaptive dynamics. This project has been supported
by grants from the National Science Foundation (NSF - Grant DMPS-0838705),
the National Security Agency (NSA - Grant H98230-09-1-0104), the Alfred
P. Sloan Foundation, the Office of the Provost of Arizona State University
and Intramural Research Program of the NIH, NCBI. This material is
also based upon work partially supported by the National Science Foundation
under Grant No. DMS-1135663.

\pagebreak


\begin{thebibliography}{9}

\bibitem{geritz1997evolutionarily} S.A.H. Geritz, E. Kisdi, G. Mesze, and J.A.J. Metz. Evolutionarily singular strategies and the adaptive growth and branching of the evolutionary tree, \emph{Evolutionary ecology}, 12(1): 35-57, 1997.

\bibitem{gorban1984equilibrium} AN Gorban, \emph{Equilibrium encircling. Equations of chemical kinetics and their thermodynamic analysis}, Nauka, Novosibirsk, 1984.

\bibitem{gorban2007selection} AN Gorban, Selection theorem for systems with inheritance, \emph{Math. Model. Nat. Phenom}, 2(4):1-45, 2007.

\bibitem{hardin2009tragedy} G. Hardin. The tragedy of the commons. \emph{Journal of Natural Resources Policy Research}, 1(3):243-253, 2009.

\bibitem{hofbauer2003evolutionary} J. Hofbauer and K. Sigmund. Evolutionary game dynamics. \emph{Bulletin of the American Mathematical Society}, 40(4):479, 2003.

\bibitem{j2005can} D. J Rankin and A. Lopez-Sepulcre. Can adaptation lead to extinction? \emph{Oikos}, 111(3):616-619, 2005.

\bibitem{karev2010mathematical} G.P. Karev. On mathematical theory of selection: continuous time population dynamics.\emph{ Journal of mathematical biology}, 60(1):107-129, 2010.

\bibitem{karev2010principle} G.P. Karev. Principle of minimum discrimination information and replica dynamics.\emph{Entropy}, 12(7):1673-1695, 2010.

\bibitem{kareva2012transitional}I. Kareva, F. Berezovskaya, and C. Castillo-Chavez. Transitional regimes as early warning signals in resource dependent competition models. \emph{Mathematical Biosciences}, 240(2): 114-123, 2012.

\bibitem{krakauer2009diversity} D.C. Krakauer, K.M. Page, and D.H. Erwin. Diversity, dilemmas, and monopolies of niche construction. \emph{The American Naturalist}, 173(1):26-40, 2009.

\bibitem{milinski2002reputation} M. Milinski, D. Semmann, and H.J. Krambeck. Reputation helps solve the tragedy of the commons.\emph{ Nature}, 415(6870):424-426, 2002.

\bibitem{miller2007complex} J.H. Miller and S.E. Page. \emph{Complex adaptive systems: An introduction to computational models of social life}. Princeton University Press, 2007.

\bibitem{nowak2006evolutionary} M.A. Nowak.\emph{ Evolutionary dynamics: exploring the equations of life}. Belknap Press, 2006.

\bibitem{ostrom1990governing} E. Ostrom. \emph{Governing the commons: The evolution of institutions for collective action}. Cambridge University Press, 1990.

\bibitem{page2008difference} S.E. Page. \emph{The Difference: How the Power of Diversity Creates Better Groups, Firms, Schools, and Societies (New Edition)}. Princeton University Press, 2008.

\bibitem{page2011diversity} S.E. Page. \emph{Diversity and complexity.} Princeton University Press, 2011.

\bibitem{vincent2005evolutionary} T.L. Vincent and J.S. Brown. \emph{Evolutionary game theory, natural selection, and Darwinian dynamics.} Cambridge University Press, 2005.

\bibitem{vollan2010cooperation} B. Vollan and E. Ostrom. Cooperation and the commons. \emph{Science}, 330(6006):923-924, 2010.

\end{thebibliography}

\newpage{}\end{doublespace}

\end{document}